\documentclass{aa}

\usepackage{array}                                              %tableaux
\usepackage{subfig}                                             %plusieurs figures avec subfloat 
\usepackage{multirow}
\usepackage{setspace}                                           
\usepackage{titletoc}
\usepackage{fancyhdr}                       % entete et pied de pages
\usepackage{multicol}                       % gestion plusieurs colonnes
\usepackage{eqnarray}                                           %maths multiligne               
\usepackage{amsmath}                            % maths avancees
\usepackage[switch, modulo]{lineno}                     %line munber
\usepackage{amssymb}
\usepackage{mathrsfs}
\usepackage{appendix}                                           % annexes
\usepackage[squaren, Gray, cdot]{SIunits}       % unites SI
\usepackage{pdflscape}                                                  %page  format paysage
\usepackage{geometry}
\usepackage{afterpage}

\usepackage{footnote}

\usepackage{color}
\usepackage{graphicx}
\graphicspath :
\graphicspath{{images/}}
\usepackage{epstopdf} 
\usepackage{txfonts}
\usepackage{booktabs} % commandes pour tableaux
\usepackage{natbib}
\usepackage{rotating}
\usepackage{xspace}
\usepackage{float}
\usepackage{url}
\usepackage[flushleft]{threeparttable}

\def\Fermi{\textit{Fermi}\xspace}
\def\gam{$\gamma$}

\begin{document}

%\linenumbers
   \title{Search for gamma-ray emission \\from superluminous supernovae with the \Fermi-LAT}
\author{
N.~Renault-Tinacci\inst{1,2} \and 
K.~Kotera\inst{1,2} \and 
A.~Neronov\inst{3} \and
S.~Ando\inst{4} 
\thanks{Corresponding authors:  n.renault.tinacci@gmail.com; kotera@iap.fr}
}
\institute{
Sorbonne Universit\'es, UPMC Univ. Paris 6 et CNRS, UMR 7095,  Institut d'Astrophysique de Paris, 98 bis bd Arago, 75014 Paris, France\goodbreak
\and
Laboratoire AIM-Paris-Saclay, CEA/DSM/IRFU, CNRS, Universite Paris Diderot,  F-91191 Gif-sur-Yvette, France
\and 
Astronomy Department, University of Geneva, Ch. d'Ecogia 16, 1290, Versoix, Switzerland\goodbreak
\and
GRAPPA Institute, University of Amsterdam, 1098 XH Amsterdam, The Netherlands\goodbreak
}
   \date{Received Month Day, Year; accepted Month Day, Year}
\abstract{ 
We present the first individual and stacking systematic search for $\gamma$-ray emission in the GeV band  in the directions of 45 superluminous supernovae (SLSNe) with the {\it Fermi} Large Area Telescope (LAT). No excess of \gam -rays from the SLSN positions was found. We report $\gamma$-ray luminosity upper limits and discuss the implication of these results on the origin of SLSNe and, in particular, the scenario of central compact object-aided SNe. From the stacking search, we derived an upper limit at 95\% confidence level (CL) to the $\gamma$-ray luminosity (above 600 MeV) $L_{\gamma}<9.1\times10^{41}$\,erg\,s$^{-1}$ for an assumed $E^{-2}$ photon spectrum for our full SLSN sample. We conclude that the rate of the neutron stars born with millisecond rotation periods {$P\lesssim 2\,$ms and $B\sim10^{12-13}\,$G} must be lower than the rate of the observed SLSNe. The luminosity limits obtained on individual sources are also constraining: in particular, \object{SN2013fc}, \object{CSS140222}, \object{SN2010kd}, and \object{PTF12dam} can only be born with millisecond periods if $B\lesssim 10^{13}\,$G. 
}
\authorrunning{Renault-Tinacci et al.}
\titlerunning{Gamma-ray detection of SLSNe}
 \maketitle

%%%%%%%%%%%%%%%%%%%%%%%%%%%%%%%%%%%%%%%%%%%%%%%%%%%
%--------------------------------------------------------%
%%%%%%%%%%%%%%%%%%%%%%%%%%%%%%%%%%%%%%%%%%%%%%%%%%%
\section{Introduction}
Superluminous supernovae (SLSNe) constitute a rare class of bright transients with luminosities ten to hundreds of times those of usual core-collapse or thermonuclear supernovae \citep{Quimby12}. With the advent of systematic transient surveys such as the  Palomar Transient Factory \citep{2009PASP..121.1334R}, Pan-STARRS1 \citep{2010SPIE.7733E..0EK}, Catalina Real-Time Transient Survey \citep{2009ApJ...696..870D}, or La Silla QUEST \citep{2013PASP..125..683B}, optical observations of a large number of these events have been collected, spanning redshifts from $z\sim 0.1$ to 4 \citep{2012Natur.491..228C}. However, the origin of these explosions is not yet understood. Mainly three scenarios have been proposed to explain these exceptionally luminous light curves, which could be i) powered by the interaction of the supernova (SN) ejecta with the circumstellar medium (e.g. \citealp{Ofek07,Quimby11,Chevalier11}), ii) pair-instability driven \citep{Gal-Yam09,Gal-Yam09b}, or iii) neutron-star driven \citep{Kasen10,Dessart12_2,KPO13,Metzger14,Murase15,Suzuki16}. Given the variety of observed spectra, it is plausible that different processes are at play in different objects (e.g. \citealp{Gal-Yam12,Nicholl14}). Interestingly, scenarios i) and iii) predict bright associated $\gamma$-ray emission in the GeV to TeV range \citep{Murase11,Murase14,Katz12,KPO13,Murase15}. The search for such \gam-ray emission with the {\it Fermi} Large Area Telescope (LAT) data is the scope of this paper. 

In the most conventional model (scenario i) SLSNe are powered by the interaction between the SN ejecta and a massive, optically thick circumstellar medium \citep{Smith07,Smith08,Miller09,Benetti14}. \object{SN2003ma} and \object{SN2006gy} for example seem to be explained well by this phenomenology \citep{Smith07,Ofek07,Smith10}. 
Several authors \citep{Murase11,Katz12,Murase14} have demonstrated that a collisionless shock could then be formed and would host efficient cosmic-ray acceleration leading to non-thermal emission from radio-submillimeter to $\gamma$-rays. In the GeV range, this emission can escape from the system without severe attenuation, for specific shock velocities (about 4500--5000 km\,s$^{-1}$) \citep{Murase15}, and at late times after the shock breakout. \cite{Fermi_CSM15} searched for this specific radiation with the LAT at the location of core-collapse SNe (Types IIn and Ib), with standard luminosity, spanning typical time windows of a few months to a year starting from the optical luminosity peak. No detection was reported and model-independent flux upper limits were derived. 

A fast-rotating central neutron star releasing its rotational energy into the SN ejecta could also drive SLSNe (model iii; e.g. \citealp{Kasen10}). The rotation period has to be close to milliseconds to transfer significant energy to the ejecta. The strength of the initial dipole magnetic field of the star sets the timescale over which the energy is injected (a stronger field leads to faster decline). Magnetars ($B\sim 10^{15}\,$G) have thus been proposed as central engines for SLSNe \citep{Kasen10,Dessart12_2}. Pulsars with millisecond periods at birth and milder dipole magnetic fields $B\sim 10^{13}\,$G would also lead to bright peaks as well as a high-luminosity plateau lasting for several months to years \citep{KPO13,Murase15}. These authors further calculated that the young neutron-star wind nebula would present a bright X-ray and $\gamma$-ray peak, respectively, through synchrotron radiation and inverse Compton (IC) scattering, appearing a few months to years after the explosion. As in model i), the $\gamma$-ray flux would be attenuated above TeV energies owing to two-photon attenuation processes, but is expected to be particularly high around $\sim 10\,$GeV \citep{Murase15}. 

We present in this work the first systematic individual and stacking search for $\gamma$-ray emission in the GeV band, with the LAT, in the directions of 45 SLSNe. We first present the SLSNe sample, dataset, and methods used to measure the \gam-ray flux from the directions of selected SLSNe through individual and stacking analyses. We report the $\gamma$-ray luminosity upper limits obtained from measurements and discuss the implication of these results on the origin of SLSNe and, in particular, the scenario of a central compact object-aided SN. \\
%%%%%%%%%%%%%%%%%%%%%%%%%%%%%%%%%%%%%%%%%%%%%%%%%%%
%--------------------------------------------------------%
%%%%%%%%%%%%%%%%%%%%%%%%%%%%%%%%%%%%%%%%%%%%%%%%%%%
\section{Superluminous supernovae sample}
\begin{figure}[tbh]
\centering %\textwidth
\includegraphics[width=\linewidth]{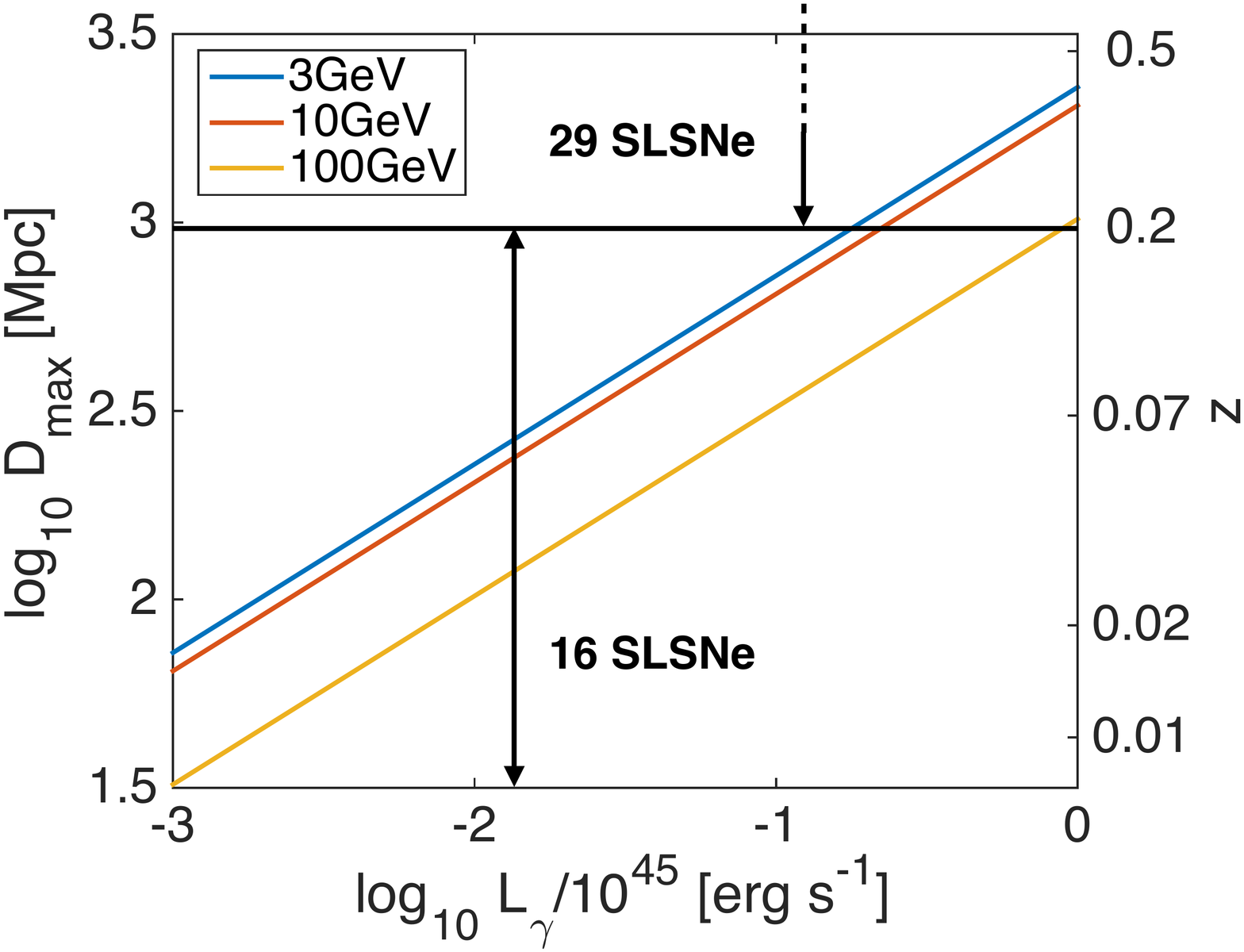}
\caption{Maximum distance $D_{\rm max}$ and corresponding redshift $z$ at which the LAT can observe a SLSN as a function of the $\gamma$-ray luminosity $L_{\gamma,\epsilon}$, for three energies:  $\epsilon=3, 10,$ and 100 GeV. The subsample boundaries are overplotted in black.}
\label{fig_lumdist_z}
\end{figure}

Superluminous supernovae reach typical optical luminosities of $\sim\,10^{42}-10^{45}$\,erg\,s$^{-1}$ \citep{Quimby11}. The \gam-ray peak luminosity could be of the same order around the peak energy $\epsilon \sim 10\,$GeV \citep{Murase15}, implying that these objects could be observed with the LAT at a given energy $\epsilon$ up to distances $D_{\rm max,\epsilon}=[{L_{\gamma,\epsilon}}/({4\pi F_{\rm{LAT,} \epsilon}})]^{1/2}$,
where $L_{\gamma,\epsilon}$ is the source $\gamma$-ray luminosity and $F_{\rm LAT, \epsilon}$ the sensitivity of the LAT\footnote{\label{foot:Fermi}  Examples of LAT Pass~8 sensitivities for 10 years:\\
$F_{\rm{LAT,} 3\, \rm{GeV}} = 1.0 \times 10^{-6} \,\,\rm{MeV}\,\rm{s}^{-1}\,\rm{cm}^{-2}$,
$F_{\rm{LAT,} 10\, \rm{GeV}} = 1.25 \times 10^{-6} \,\, \rm{MeV}\,\rm{s}^{-1}\,\rm{cm}^{-2}$,
$F_{\rm{LAT,} 100\, \rm{GeV}} = 5.0 \times 10^{-6}\,\,\rm{MeV}\,\rm{s}^{-1}\,\rm{cm}^{-2} $ \\
obtained from \url{http://www.slac.stanford.edu/exp/glast/groups/canda/lat_Performance.htm}.}  for Pass8 \citep{2013arXiv1303.3514A}, which are both calculated at energy $\epsilon$. 
In particular, 
\begin{eqnarray}\label{eq:Dmax}
%D_{\rm max,\epsilon}&=&\left(\frac{L_{\gamma,\epsilon}}{4\pi F_{\rm{Fermi,} \epsilon}}\right)^{1/2}\\
D_{\rm max,10\,GeV}\sim  2.0\times 10^3 \, {\rm Mpc}\,\, \left(\frac{L_{\rm \gamma,10\,GeV}}{10^{45}\,{\rm erg\,s^{-1}}}\right)^{1/2} \ .%\left(\frac{F_{\rm{Fermi,} \epsilon}}{F_{\rm{Fermi, 10\, \rm{GeV}}}}\right)^{-1/2} \, ,
\end{eqnarray}The above luminosity distance corresponds to a redshift $z\sim 0.36$. We selected 45 SLSNe listed in Tab.~\ref{tab_studied_SNe} among which 25 are located below this limit. Figure~\ref{fig_lumdist_z} presents the maximum observable distances with the LAT as a function of the source luminosity for different energies $\epsilon$. 

In principle, as discussed in section~\ref{sec:discussion}, scenarios i) and iii) predict $\gamma$-ray luminosities that are at least a factor 1/20 to 1/15 lower than the bolometric radiated luminosity. It thus seems more reasonable to look for sources with maximum luminosities of order $L_\gamma \sim 10^{44}\,$erg\,s$^{-1}$ within redshifts $z\lesssim 0.2$. For completeness, we still include the more distant sources in our systematic search. We consider a full sample and two subpopulations bounded by the redshift values  0.0, 0.2, and 1.6.

%%%%%%%%%%%%%%%%%%%%%%%%%%%%%%%%%%%%%%%%%%%%%%%%%%%
%--------------------------------------------------------%
%%%%%%%%%%%%%%%%%%%%%%%%%%%%%%%%%%%%%%%%%%%%%%%%%%%
\section{\Fermi-LAT observations}
%Fermi LAT
The LAT, the main instrument on the \Fermi spacecraft, is a pair-conversion telescope that is sensitive to \gam-rays from 20 MeV to $>300$ GeV with on-axis effective area $> 1$ GeV of $\sim8000$ cm$^2$.%, exceeding that of EGRET \citep{1993ApJS...86..629T} by a factor of about 5. 
The LAT is made of a high-resolution silicon tracker, a hodoscopic CsI electromagnetic calorimeter and an anti-coincidence detector for charged particle background identification. The full description of the instrument and its performance can be found in \citet{LATinstrument}. The LAT field of view ($\sim2.4$ sr) covers the entire sky every 3 hr (two orbits) in the survey mode used for this work. The single-event point spread function (PSF) strongly depends on both the energy and conversion point in the tracker, but less on the incidence angle. For 1 GeV normal incidence conversions in the upper section of the tracker the PSF 68\% containment radius is $0.6^{\circ}$. Timing is provided to the LAT by the satellite GPS clock and photons are timestamped to an accuracy better than 300 ns. The photons detected by the LAT are categorized in classes according to the energy, direction reconstruction quality, and residual background rates. The categories have different respective strengths depending on the type of analysis (transient, point source, and diffuse emission).

%Data
 We used for our analysis the Pass 8 LAT data \citep{2013arXiv1303.3514A}, which was collected starting 2008 August 4 and extending until 2015 September 10. This dataset encompasses seven years and one week of observations. There are six main classes within the Pass 8 event reconstruction strategy with the classes nested\footnote{Description of the Pass 8 classes at: \url{https://fermi.gsfc.nasa.gov/ssc/data/analysis/documentation/Cicerone/Cicerone_Data/LAT_DP.html#PhotonClassification}}. We selected photons from the \textquotedblleft Source\textquotedblright ~class, which is the third event set in terms of residual charged-particle background and is mainly dedicated to the study of point sources. We kept photons within a radius of 16$^{\circ}$ from the source position and excluded the periods when the source was viewed at zenith angles $> 100^{\circ}$ to minimize contamination by photons generated by cosmic-ray interactions in the atmosphere of the Earth. Only photons within the energy range of $600$ MeV to $600$ GeV were selected. 
We performed  the analyses in seven energy bands between $0.612\,$GeV and $600\,$GeV and in the full energy range. Table~\ref{tab_analysis_eny_band} reports the energy bands used. 
The energy boundaries are determined by the energy binning of the LAT Collaboration diffuse model\footnote{http://fermi.gsfc.nasa.gov/ssc/data/access/lat/BackgroundModels.html} \citep{2016ApJS..223...26A}. We chose to use the same binning in our analysis.
However to increase photon statistics in particular at the highest energies, we merged the diffuse model energy bands to obtain the larger energy ranges used in our study. 
Theoretical models \citep{Murase15} predict a rather flat spectrum that is consistent with a spectral index of $\sim-2$ and owing to the poor angular resolution at low energy ($50-600\,$MeV), we do not expect photons at those energies to contribute significantly to our sensitivity.

\begin{table}[h]
\centering
\caption{Energy band boundaries.} % for the 7+1 and the 3+1 analyses.}
\label{tab_analysis_eny_band}
\begin{tabular}{|c|c|c|c|c}
\hline
$E_{\rm min}$ & $E_{\rm max}$ \\
%\noalign{\smallskip}
\footnotesize{[GeV]} & \footnotesize{[GeV]} \\
\hline
%\noalign{\smallskip} 
0.612 & 1.566 \\
%\noalign{\smallskip} 
1.566 & 4.005 \\
%\noalign{\smallskip} 
4.005 & 10.245 \\ 
%\noalign{\smallskip} 
10.245 & 26.207 \\ 
%\noalign{\smallskip} 
26.207 & 67.041 \\ 
%\noalign{\smallskip} 
67.041 & 171.500 \\ 
%\noalign{\smallskip} 
171.500 & 600.000 \\ 
%\noalign{\smallskip}
0.612 & 600.000 \\
\hline
\end{tabular}
\end{table}
%%%%%%%%%%%%%%%%%%%%%%%%%%%%%%%%%%%%%%%%%%%%%%%%%%%
%--------------------------------------------------------%
%%%%%%%%%%%%%%%%%%%%%%%%%%%%%%%%%%%%%%%%%%%%%%%%%%%
\section{Analysis}
\subsection{Analysis by maximum likelihood estimator (MLE)}

\subsubsection{Individual analysis}
\label{sec_indiv_ana}

The analysis method is similar to that described in Renault-Tinacci et al. (a\&b, 2 papers in prep.; \citeyear{2015ICRC...34..843R}). We describe the main steps below. 
The spectral analysis was carried out in the energy bands listed in Tab~\ref{tab_analysis_eny_band}. We test both wide and narrower energy bands to observe the impact of the increase of photon statistics and of a better sampling of the source spectrum.
We modelled the \gam-ray emission in an $18^{\circ}\times18^{\circ}$ square region centred on the position of each source. The model consists of a linear combination of a point source at the SN position and of template maps for the diffuse interstellar emission and the isotropic flux resulting from the extragalactic \gam-ray background and residuals due to cosmic rays misclassified as \gam-rays. The interstellar component and isotropic spectra are available at the \textit{Fermi} Science Support Centre\footnote{\url{http://fermi.gsfc.nasa.gov/ssc/data/access/lat/BackgroundModels.html}}. The model also includes all point sources and extended sources listed in the 3FGL catalogue \citep{3FGL}. 

The $\gamma$-ray intensity in each $(l,b)$ direction in Galactic coordinates, $I(l,b,E)$ in cm$^{-2}$~s$^{-1}$~sr$^{-1}$~MeV$^{-1}$, is modelled at each energy $E$ as
\begin{equation}
   \begin{array}{ll}
      I(l,b,E) = & S_{\rm SN}(E) \, \delta(l-l_{\rm SN},b-b_{\rm SN}) + q_{\rm{ISM}}(E) \, I_{\rm{ISM}}(l,b,E) + \\
                   & q_{\rm iso}(E) \, I_{\rm iso}(E) + \sum_j q_{S_j}(E) \, S_j(E) \, \delta(l-l_j,b-b_j) + \\
                  &  q_{S{\rm ext}}(E) \, S_{\rm ext}(l,b,E)
   \end{array}
   \label{equa_mdl_gam}
,\end{equation} 
where $S_{\rm SN}(E)$ gives the source spectrum in cm$^{-2}$~s$^{-1}$~MeV$^{-1}$ and the $I_{\rm{ISM}}(l,b,E)$ and $I_{\rm{iso}}(E)$ terms denote the interstellar and isotropic intensities in cm$^{-2}$~s$^{-1}$~sr$^{-1}$~MeV$^{-1}$, respectively.  The $q_{\rm{ISM}}(E)$ and $q_{\rm iso}(E)$ parameters are simple normalization factors to account for possible deviations from the two input spectral shapes. 

Depending on the latitude of the analysis region, the number of background sources in the region varied from 7 to 27 with an average number around 12. We used the source flux spectra $S_j(E)$ from the catalogue as input spectra for the sources (in cm$^{-2}$~s$^{-1}$~MeV$^{-1}$). Their individual flux normalizations $q_{S_j}(E)$ have been let free in each energy band to compensate for potential deviations between the four-year long observations of the catalogue and our extended dataset.

We modelled the $\gamma$-ray intensity inside the analysis region and in a $7\degree$-wide peripheral band to account for photons spilling over inside the analysis region because of the wide LAT PSF. The contribution from the sources detected in the outer band has been summed into a single map $S_{\rm ext}(l,b,E)$ and its normalization $q_{S_{\rm ext}}(E)$ has been left as a free parameter \citep{2015A&A...582A..31P}.

\begin{figure}[tbh]
\centering %\textwidth
\includegraphics[width=\linewidth]{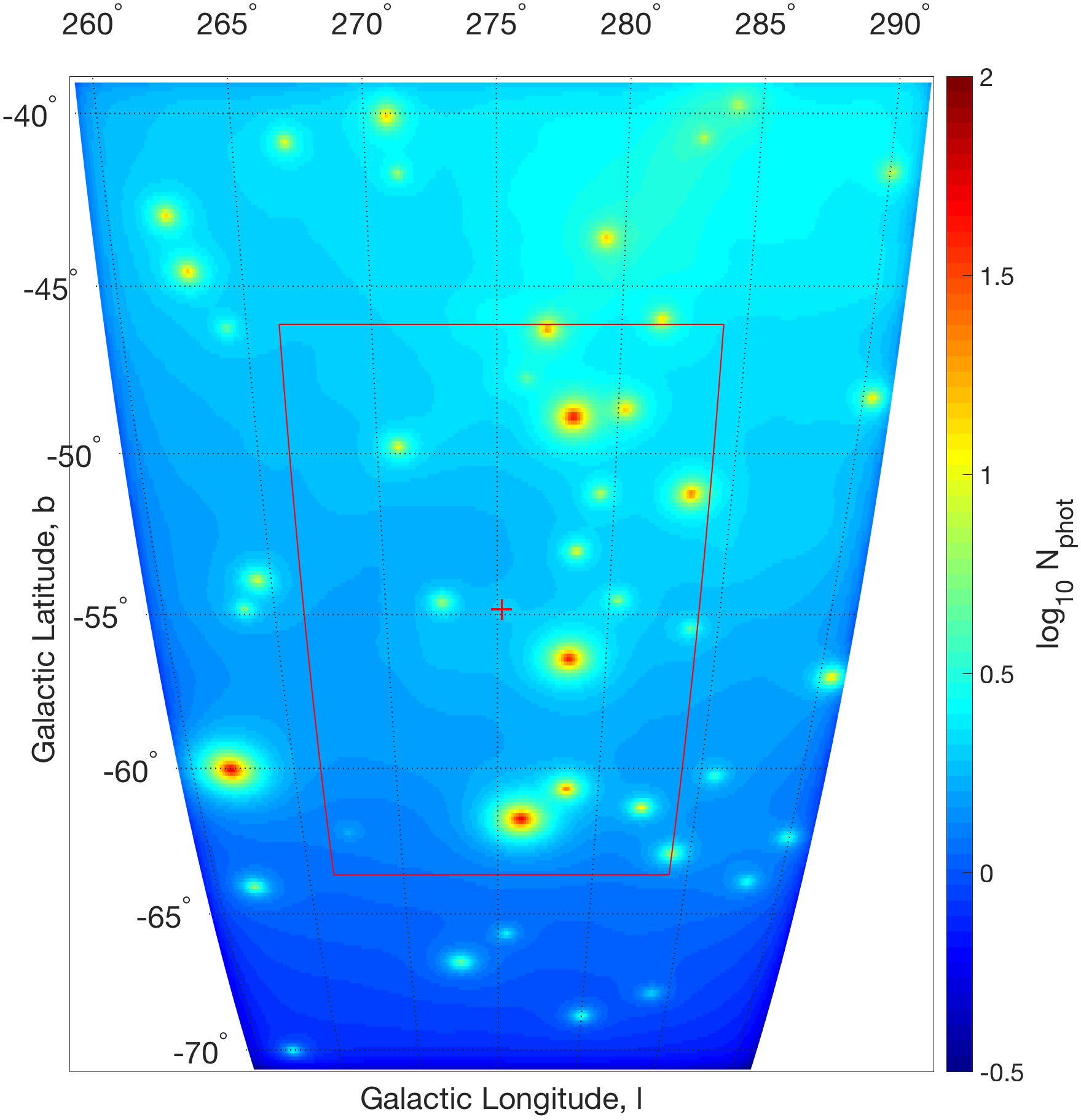} 
\caption{Source model map of \object{SN2013fc} after fit integrated over the 600\,MeV--10\,GeV energy band. The red cross points to the position of the \object{SN2013fc}. The thin red line indicates the boundary between the inner map where sources are independently fitted and the peripheral band where the contribution from all sources is fitted as one. All other sources, included in the source model map were previously detected with the \Fermi-LAT and their fluxes were fitted during the procedure.}
\label{fig_mdl_skymap}
\end{figure}

The model intensity $I(l,b,E)$ has been processed through the LAT instrument response functions (IRFs, P8R2\_V6SOURCE) to take into account the position-dependent and energy-dependent exposure on the sky and the energy-dependent PSF.  
We calculated the effective IRFs for the spectrum of each component, taking a power-law spectrum with a photon index equal to -2 as spectral input for the studied SLSNe. 
An example of a model sky map is presented on Fig.~\ref{fig_mdl_skymap} for \object{SN2013fc}.

The modelled photon map, integrated over each energy band, can be compared to the observed data by means of a binned maximum-likelihood estimator with Poisson statistics to fit the model coefficients to the LAT data \citep{2015arXiv150203081A}. We stress that the present analysis independently fits the source flux in each energy band and that it is independent of the initially assumed spectral shape. 

To quantify the detection significance of the emission, we used the Test Statistic, $TS = 2 \left[ln(L)-ln(L_0)\right]$, where $L_0$ and $L$ are the maximum-likelihood values obtained for the null hypothesis (zero flux from the SN) and when a point source at SN position is assumed, respectively. In the first (and fairly robust) approximation in case of a difference of 1\,degree of freedom, the significance is equal to the square root of the TS. We set a detection threshold above a TS of 25, i.e. a 5$\sigma$ significance.

Performing multiple analysis trials with different parameters, as we have here with several SLSNe for instance, introduces a bias in the analysis due to the so-called look-elsewhere effect \citep{2011arXiv1101.0390C}. The chance that the observed significance could have arisen at random due to the size of the parameter space that was searched can be accounted for by applying trials factor corrections to the final significance. For the individual search study, we have to consider 45 SLSNe, seven individual and independent energy bands, the total energy range that overlaps the individual bands, and five overlapping time windows. This corresponds to a number of trials $N_{\rm trials}=1800$. For such a high number of trials, a 5$\sigma$ (4$\sigma$) pre-trial detection would correspond to a 3.3$\sigma$ (1.6$\sigma$) significance after trials correction. The best significance, obtained for \object{SN2012il} in the three-month time window and the 67-172 GeV band, is equal to 3.8$\sigma$, which is decreased to 1.2$\sigma$ after applying the trials factor. This correction assumes that each dataset is statistically independent, which is overly conservative in our case, since four of the five time windows result in datasets that are subsets of each other and the total energy range overlaps the individual energy bands. Therefore we regard the final detection significances as conservative lower limits to the true significances of the signal over the background.

We checked the existence of a steady \gam-ray source in the direction of the selected SLSNe. Such a source could be the host galaxy of the SN or a source along the line of sight.
 In case of detection (TS$>25$, significance$> 5\sigma$), a new source would be added to the catalogued sources at the position of the SLSN. 
 We searched for \gam-ray emission in the SLSN off-peak dataset, i.e. either between the first available \gam-ray observations (2008 August 8) and one month before the presumed date of the SN peak time, $t_{\rm SN}$, or for three SLSNe (\object{SN2008fz}, \object{SN2009jh}, and \object{PTF09atu}) the two last years before the end of the exploited dataset (2015 September 10) because their peak time is less than one year after the observations start. 
 No significant source was found in the SN directions in the off-peak window.
 The existence of significant \gam-ray emission from the host galaxy or an aligned source would make the detection of a faint SN signal more difficult.\\

Theoretical simulations of the duration of the \gam-ray emission predict \gam-ray emission lasting weeks to months depending on the SN and the central compact object characteristics \citep{Murase15}. Hence this prediction motivates a search in several time windows. Individual and joint likelihood analyses were performed for the following observation periods:
\begin{itemize}
\item from $t_{\rm peak}$\footnote{If peak time is not known, detection time is used.} $-1$ month (referred to as $t_{\rm SN}$ in the next sections) up to $t_{\rm peak}+3$ months
\item from $t_{\rm peak}-1$ month up to $t_{\rm peak}+6$ months
\item from $t_{\rm peak}-1$ month up to $t_{\rm peak}+1$ year
\item from $t_{\rm peak}-1$ month up to $t_{\rm peak}+2$ years.
\end{itemize}
Only 44, 40, and 33 SNe were studied for the six-month, two- and one-year time windows, respectively, because the available dataset was too short for the needed duration. On the other hand, all sources were analysed for the three-month time windows.
To make sure no early \gam-ray emission was missed, we used datasets starting 30 days before the optical peak time to account for the uncertainty in its determination \citep{2015MNRAS.452.1535C,2016arXiv161207321L}.

\subsubsection{Joint likelihood analysis}

To improve the sensitivity of the analysis to a weak \gam-ray signal from SLSNe, we combined sources in a joint likelihood analysis \citep{2015arXiv150203081A}. We studied the complete sample and split it into two sub-groups based on redshift (and hence also on distance). Fig.~\ref{fig_lumdist_z} summarizes the repartition of studied SLSNe. Some sources exploded late with respect to the dataset time limits preventing us from including these sources into the joint likelihood analysis for the longer time bins. Only the three-month analysis includes the complete sample. Otherwise, we used sources from \object{SN2008fz} to \object{DES13S2cmm}, \object{SN2008fz} to \object{PS1-14bj}, and \object{SN2008fz} to \object{SN2015bn} for the two-year, one-year, and six-month
analyses, respectively (following the order in Table~\ref{tab_studied_SNe}).

 To be independent from any spectral shape assumption we performed the analysis in energy bands (see Section~\ref{sec_indiv_ana} for details). We performed the combined analysis by tying together in each energy band the flux normalization of all SLSNe in the subsample \citep{Fermi_CSM15}. It results in a single free parameter per energy band. To correctly tie the SN normalizations together, we defined a common \gam-ray scaling factor; i.e. we give more weight to sources with greater expected \gam-ray flux in the joint likelihood. Two different weighting approaches can be envisaged in the stacking procedure, relying either on the optical flux or the luminosity distance. Considering the difficulties in concatenating a consistent set of optical magnitude values for the whole sample, we ruled out the optical flux approach. For the joint analysis we assumed all SLSNe to have the same intrinsic \gam-ray luminosity and thus the observed \gam-ray flux of each source scales with a factor inversely proportional to the luminosity-distance squared. The weight of the flux normalization of each source in each energy bin is $w_{d} = (100\,\rm{Mpc}/\rm{d})^2$.

We derived the SLSNe distances from redshift measurements and a set of cosmological parameters from the $\Lambda$CDM model. We used $H_0\,=\,69.6\,$km~s$^{-1}$~Mpc$^{-1}$, $\Omega_M\,=\,0.286$, and $\Omega_{\Lambda}\,=\,0.714$ values provided in \citet{Planck2015_cosmoparam} but other sets exist. Hence we calculated roughly that the choice of a different set would result in distance estimates less than 5\% greater or lower with other commonly used cosmological parameters \citep[e.g.][]{Planck2013_cosmoparam,2013ApJS..208...19H,Nicholl14,Benetti14}.

For the joint likelihood analysis, only the SN normalization is free in each energy band while the nearby source, diffuse, and isotropic component normalizations are fixed to their values obtained in the individual analyses of each source in the corresponding energy band.

Identically to the individual searches, many trials are realized for the joint likelihood analysis. We count here four time windows, seven small energy bands and the total range, and two redshift subpopulations and the full sample, which brings us to $N_{\rm trials} = 96$. Again the correction is conservative as the redshift subsets overlap along with the time windows and total energy range. In this case,  a 5$\sigma$, 4$\sigma,$ or 3$\sigma$ significance would correspond to a 4$\sigma$, 2.7$\sigma,$ or 1.2$\sigma$ post-trial detection level, respectively. With the total SLSN population, the two-year time window and in the 67-172 GeV band, a 3.8$\sigma$ significance is obtained and is decreased to 2.4$\sigma$ after trials factor correction. We thus report only upper limits.

\subsection{Aperture photometry}

We independently verified the results of the likelihood analysis using the aperture photometry method for spectral extraction. For the aperture photometry, we extracted the source signal from circular regions of radius $1^\circ$ around each source listed in Table \ref{tab_studied_SNe}. Photons of the "Source" class were retained for the analysis. For each source region the exposure was calculated using the {\tt gtexposure} tool, which accounts for the energy-dependent loss of the source signal due to the large size of the LAT point spread function extending beyond the $1^\circ$ around the source position energies.

The background was estimated from source-free regions of radius $3^\circ$ within $<10^\circ$ distance from the source position. This assures that the level of the Galactic diffuse background in the source and background estimate regions is similar. The level of the Galactic diffuse background varies on different angular scales. This is the main limitation of the aperture photometry method, especially for the sources close to the Galactic Plane. However, most of the sources considered for the stacking analysis are at high Galactic latitudes where the level of variations of the Galactic diffuse emission is more moderate and their angular scale is typically larger than a few degrees. This justifies the use of the aperture photometry as a cross-checking method.

The two analysis methods are complementary in the sense that the aperture photometry provides a robust upper limit on the luminosity, which is independent of the details of modelling of diffuse backgrounds in the source region of interest. At the same time, the (moderately) model-dependent likelihood analysis allows us to tighten the upper limits on the luminosity of the SLSN source sample. 

%%%%%%%%%%%%%%%%%%%%%%%%%%%%%%%%%%%%%%%%%%%%%%%%%%%
%--------------------------------------------------------%
%%%%%%%%%%%%%%%%%%%%%%%%%%%%%%%%%%%%%%%%%%%%%%%%%%%
\section{Results}
In the following section, we report the upper limits at $2\sigma$ confidence on the summed luminosities, $L^{\rm sum}_{0.6-600.0\,\rm{GeV}}$ and $L^{\rm sum}_{1-10\,\rm{GeV}}$. These are the sum of measured luminosities in the individual energy bands located between the indicated energy boundaries. 
These luminosities are named this way in contrast to the total luminosity obtained directly by fitting the flux in the studied energy band. On the other hand, the summed luminosity is the sum of the fluxes obtained by fitting fluxes separately in narrow energy bins covering the large energy band and summing them afterwards. The second method via summation allows us to reach a better sampling of the actual source spectrum compared to the first method through a direct fit that provides a rougher measure of the luminosity.
To compute the upper limits on the \gam-ray luminosity in individual energy bands from the measured photon fluxes and their errors, we assumed, as for the input spectrum, a $E^{-2}$ power law in each energy band. We emphasize that the derived upper limits on $L^{\rm sum}_{1-10\,\rm{GeV}}$ are the most important measurements to probe the theoretical predictions.

\subsection{Individual analysis}

\begin{figure}[tbh]
%\centering %\textwidth
\includegraphics[width=\linewidth,trim={10mm 8mm 21mm 20mm},clip]{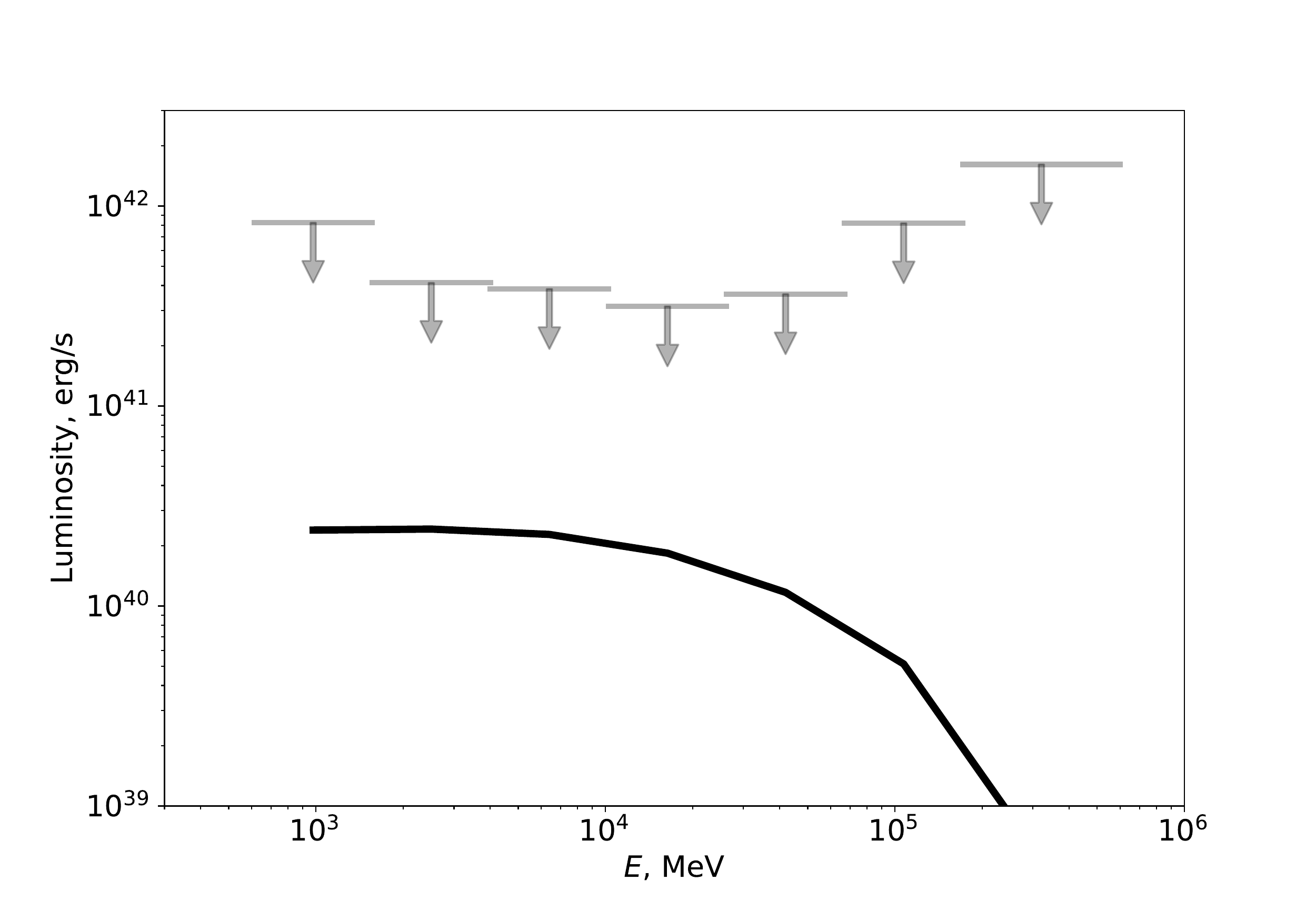}
\caption{Upper limits on the luminosity of \object{SN2013fc} from the individual analysis of the 2-year time window. Down arrows indicate upper limits at $2\sigma$ confidence. The black line represents the integrated luminosity computed from a simulated spectrum derived by \cite{Murase15} for a neutron star with $P=10\,$ms, $B=10^{13}\,$G at a distance $d=16.5\,$kpc and about 206 days after explosion.}
\label{fig_SN2013fc_2y_7bnd}
\end{figure}

We gather in Tab\ref{tab_SNe_fluxes_3m-2y_7bnd} the individual luminosities or upper limits at $2\sigma$ confidence. No detection, after trial factor correction, are reported over the total energy range nor in individual bands. The only obtained small over-fluctuaction occurs in an individual energy band for \object{SN2012il} between 67 and 172 GeV at 1.2, 0.9, 0.7, and 0.3 $\sigma$ levels in the three-month to two-year time windows, respectively. It is however insignificant. The most constraining upper limit for an individual source was obtained for \object{SN2013fc} with a two-year time window and is equal to $1.2\times10^{42}$\,erg\,s$^{-1}$ (see Fig.~\ref{fig_SN2013fc_2y_7bnd}).

\subsection{Joint likelihood analysis}

\begin{table*}[!h]
\centering 
\caption{Luminosities from joint likelihood analysis measurements with all sources. The first and second columns contain the upper limits on the sum of the derived luminosities in the individual energy bands between 600 MeV and 10 GeV, and 600 MeV and 600 GeV, respectively. The third column indicates the highest post-trial significance in an individual energy band and the two last columns report the correspond energy band boundaries.}
\label{tab_SNe_fluxes_stacking_all_src_7bnd} 
\begin{tabular}{lcccccccc} 
Time window & $L_{0.6-10.2\,\rm{GeV}}$ & $L_{0.6-600.0\,\rm{GeV}}$ & $\rm{Sig}_{E1-E2}^{\rm best\,bnd}$ & E1 & E2 \\ 
\noalign{\smallskip} 
 & [erg\,s$^{-1}$] & [erg\,s$^{-1}$] & [$\sigma$ units] & [GeV] & [GeV] \\ 
\hline 
\noalign{\smallskip} 
$t_{\rm SN}$ to $t_{\rm SN}$ $+\,3$ months & $<6.2\times10^{42}$ & $<1.8\times10^{44}$ & 1.2 & 171.50 & 600.00 \\ 
\noalign{\smallskip} 
$t_{\rm SN}$ to $t_{\rm SN}$ $+\,6$ months & $<3.1\times10^{42}$ & $<1.0\times10^{44}$ & 0.9 & 171.50 & 600.00 \\ 
\noalign{\smallskip} 
$t_{\rm SN}$ to $t_{\rm SN}$ $+\,1$ year & $<1.3\times10^{42}$ & $<2.3\times10^{43}$ & 0.0 & 67.04 & 171.50 \\ 
\noalign{\smallskip} 
$t_{\rm SN}$ to $t_{\rm SN}$ $+\,2$ years & $<9.1\times10^{41}$ & $<2.4\times10^{43}$ & 2.4 & 67.04 & 171.50 \\ 
\noalign{\smallskip} 
SN off-peak period & $<6.0\times10^{41}$ & $<3.8\times10^{42}$ & 0.1 & 26.21 & 67.04 \\ 
\noalign{\smallskip} 
\hline 
\end{tabular} 
%\end{table*} 
%
\vspace{0.5cm}
%\begin{table*}
\centering 
\caption{Luminosities from joint likelihood analysis measurements and sources with redshift between 0.0 and 0.2. The layout of the table is identical to Table~\ref{tab_SNe_fluxes_stacking_all_src_7bnd}.} 
\label{tab_SNe_fluxes_stacking_redshift1_7bnd} 
\begin{tabular}{lcccccccc} 
Time window & $L_{0.6-10.2\,\rm{GeV}}$ & $L_{0.6-600.0\,\rm{GeV}}$ & $\rm{Sig}_{E1-E2}^{\rm best\,bnd}$ & E1 & E2 \\ 
\noalign{\smallskip} 
 & [erg\,s$^{-1}$] & [erg\,s$^{-1}$] & [$\sigma$ units] & [GeV] & [GeV] \\ 
\hline 
\noalign{\smallskip} 
$t_{\rm SN}$ to $t_{\rm SN}$ $+\,3$ months & $<7.1\times10^{43}$ & $<8.9\times10^{44}$ & 0.5 & 67.04 & 171.50 \\ 
\noalign{\smallskip} 
$t_{\rm SN}$ to $t_{\rm SN}$ $+\,6$ months & $<3.2\times10^{43}$ & $<5.5\times10^{44}$ & 0.3 & 67.04 & 171.50 \\ 
\noalign{\smallskip} 
$t_{\rm SN}$ to $t_{\rm SN}$ $+\,1$ year & $<1.7\times10^{43}$ & $<3.1\times10^{44}$ & 0.2 & 67.04 & 171.50 \\ 
\noalign{\smallskip} 
$t_{\rm SN}$ to $t_{\rm SN}$ $+\,2$ years & $<1.1\times10^{43}$ & $<2.9\times10^{44}$ & 1.0 & 171.50 & 600.00 \\ 
\noalign{\smallskip} 
SN off-peak period & $<1.4\times10^{43}$ & $<9.8\times10^{43}$ & 0.0 & 1.57 & 4.00 \\ 
\noalign{\smallskip} 
\hline 
\end{tabular} 
%\end{table*} 
%
\vspace{0.5cm}
%\begin{table*} 
\centering 
\caption{Luminosities from joint likelihood analysis measurements and sources with redshift between 0.2 and 1.6. The layout of the table is identical to Table~\ref{tab_SNe_fluxes_stacking_all_src_7bnd}.} 
\label{tab_SNe_fluxes_stacking_redshift5_7bnd} 
\begin{tabular}{lcccccccc} 
Time window & $L_{0.6-10.2\,\rm{GeV}}$ & $L_{0.6-600.0\,\rm{GeV}}$ & $\rm{Sig}_{E1-E2}^{\rm best\,bnd}$ & E1 & E2 \\ 
\noalign{\smallskip} 
 & [erg\,s$^{-1}$] & [erg\,s$^{-1}$] & [$\sigma$ units] & [GeV] & [GeV] \\ 
\hline 
\noalign{\smallskip} 
$t_{\rm SN}$ to $t_{\rm SN}$ $+\,3$ months & $<4.1\times10^{43}$ & $<8.9\times10^{44}$ & 0.0 & 171.50 & 600.00 \\ 
\noalign{\smallskip} 
$t_{\rm SN}$ to $t_{\rm SN}$ $+\,6$ months & $<2.3\times10^{43}$ & $<5.2\times10^{44}$ & 0.0 & 171.50 & 600.00 \\ 
\noalign{\smallskip} 
$t_{\rm SN}$ to $t_{\rm SN}$ $+\,1$ year & $<1.5\times10^{43}$ & $<1.7\times10^{44}$ & 0.0 & 1.57 & 4.00 \\ 
\noalign{\smallskip} 
$t_{\rm SN}$ to $t_{\rm SN}$ $+\,2$ years & $<1.1\times10^{43}$ & $<1.1\times10^{44}$ & 0.9 & 67.04 & 171.50 \\ 
\noalign{\smallskip} 
SN off-peak period & $<5.9\times10^{42}$ & $<4.7\times10^{43}$ & 0.0 & 26.21 & 67.04 \\ 
\noalign{\smallskip} 
\hline 
\end{tabular} 
\end{table*} 

\begin{figure}[tbh]
%\centering 
%\textwidth
\includegraphics[width=\linewidth,trim={8mm 1mm 23mm 13mm},clip]{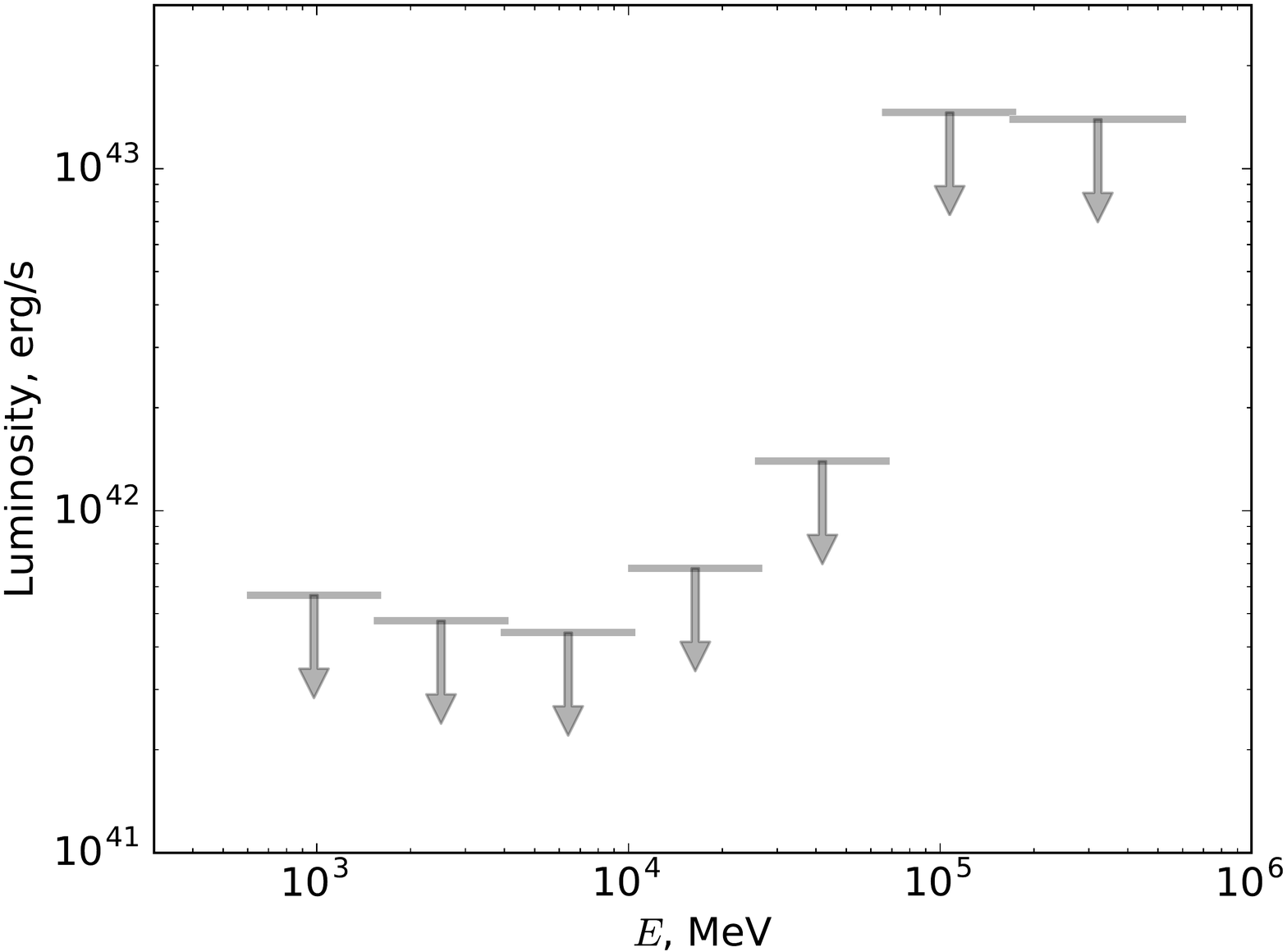}
\caption{Upper limits on the luminosity of the SLSN from joint likelihood analysis for a 2-year time window and a sample containing all sources (Table~\ref{tab_SNe_fluxes_stacking_all_src_7bnd}). Down arrows indicate upper limits at $2\sigma$.}
\label{fig_stacking_2y_7bnd}
\end{figure}

Tables~\ref{tab_SNe_fluxes_stacking_all_src_7bnd} to \ref{tab_SNe_fluxes_stacking_redshift5_7bnd} provide the upper limits on luminosity for the stacking analysis obtained for both energy band sets. The study was carried out with three different source samples:
\begin{itemize}
\item all sources
\item sources with redshift $z \in\,[0.0\,;\,0.2]$
\item sources with redshift $z \in\;]0.2\,;\,1.6]$.
\end{itemize}
Joint likelihood fits result, after trial factor correction, for either of the redshift samples and time windows, in no detection. A 2.4$\sigma$ over-fluctuation (post trials) can be reported in the 67-172 GeV individual energy band with the two-year dataset and the complete sample. The latter includes mostly ($\sim66\%$) sources that are too distant to be detectable individually and hence we considered this fluctuation as insignificant. Joint likelihood analyses on the subsample of lowest redshifts (closest SLSNe) should be more relevant if any signal was detected. In the absence of signal, i.e. when we are looking at an empty sky, the tightest upper limit is provided by the joint analysis with the largest sample and the longest dataset and is independent of the source distances.

As expected since we do not detect anything but an over-fluctuation, the most constraining limit on the luminosity, $L_{\gamma}<9.1\times10^{41}\,$erg\,s$^{-1}$, is obtained for the total population of SLSNe and the two-year time window. Figure~\ref{fig_stacking_2y_7bnd} represents the luminosity upper limits as a function of energy for this case. We obtained upper limits with the subsample for z>0.2 lower than for z<0.2 and this is simply explained by the fact that the subset z>0.2 is larger than the subpopulation with z<0.2 and both samples result in no detection and hence are basically two empty skies. We considered both close and farthest SLSNe because we simply took into account all the catalogued SNe. Nonetheless we had in mind that those with redshift beyond 0.2 would be theoretically undetectable.

\begin{figure} %[h] %
%\centering
\includegraphics[width=\linewidth,trim={4mm 1mm 16mm 15mm},clip]{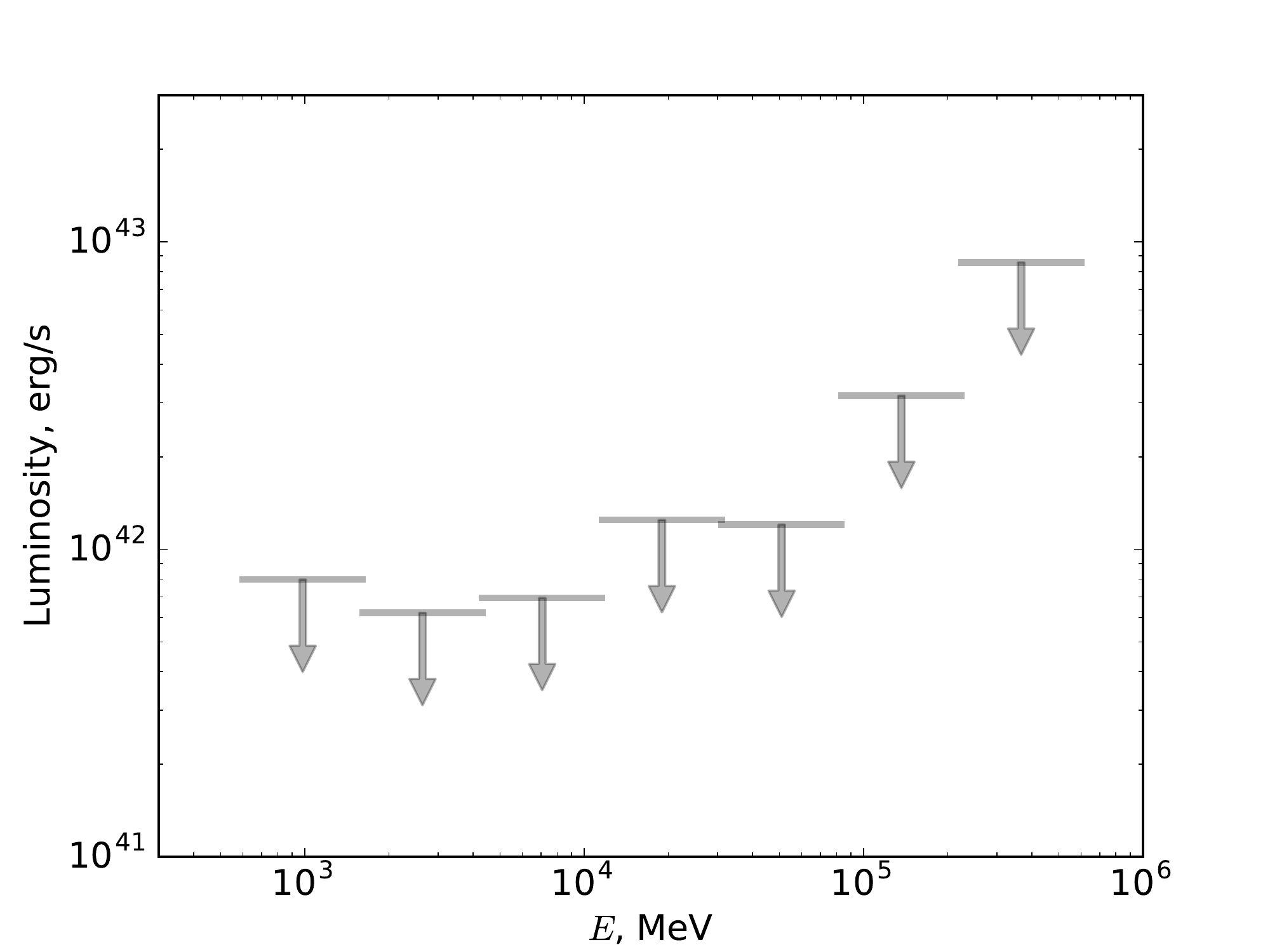}
\caption{Upper limits on the luminosity of the SLSN obtained from the aperture photometry for a 2-year time window and a sample containing all sources. Down arrows indicate upper limits at $2\sigma$.}
\label{fig:aperture}
\end{figure}

The results of the aperture photometry analysis, shown in Fig. \ref{fig:aperture} for the two-year time span, are consistent with those of the likelihood analysis, although the upper limits on the luminosity of the stacked source sample are somewhat higher (by $\simeq 30\%$) in the GeV energy band. At the same time, in the energy range above 100 GeV the aperture photometry bound is tighter than that derived from the likelihood analysis. This is explained by the fact that the background photon statistics in this energy band are low and the signal is detectable in a nearly background-free regime. Modelling of the diffuse background in the likelihood analysis approach introduces additional parameters in the analysis and, as a consequence, slightly relaxes the bounds on the source flux and luminosity. 
However, modelling the diffuse background precisely is more critical at low energy and for sources close to the Galactic plane. Hence, modifying the diffuse model in this analysis ($E>600\,$MeV and high latitudes) would only have a tiny effect. \citet{1st_Fermi_SNR_cat} performed a study of the systematic errors due to the choice of the diffuse model.

%%%%%%%%%%%%%%%
%%%%%%%%%%%
%%%%%%%%%%%%%%%%%%%%%%%%%%%%%%%%%%%%%%%%%%%%%%%%%%%
%--------------------------------------------------------%
%%%%%%%%%%%%%%%%%%%%%%%%%%%%%%%%%%%%%%%%%%%%%%%%%%%
\section{Discussion}
\label{sec:discussion}

We discuss here the implication of the derived contraints on the \gam-ray luminosity received from SLSNe in the scenario of central compact object-aided SN.
In this scenario, a fast-rotating central neutron star releases its rotational energy into the SN ejecta via its wind \citep[see introduction and see e.g.][]{Kasen10,KPO13}. The electromagnetic energy of the pulsar is dissipated into kinetic energy in the wind at a yet unknown location (see e.g. Kirk et al. 2009), seeding the surrounding pulsar wind nebula with accelerated pairs. These are expected to radiate by synchrotron or IC scattering and possibly produce a \gam-ray emission. The simulations presented in \cite{Murase15} in this framework were used as a basis for the discussion.

\begin{figure}
\centering %\textwidth
\includegraphics[width=\linewidth]{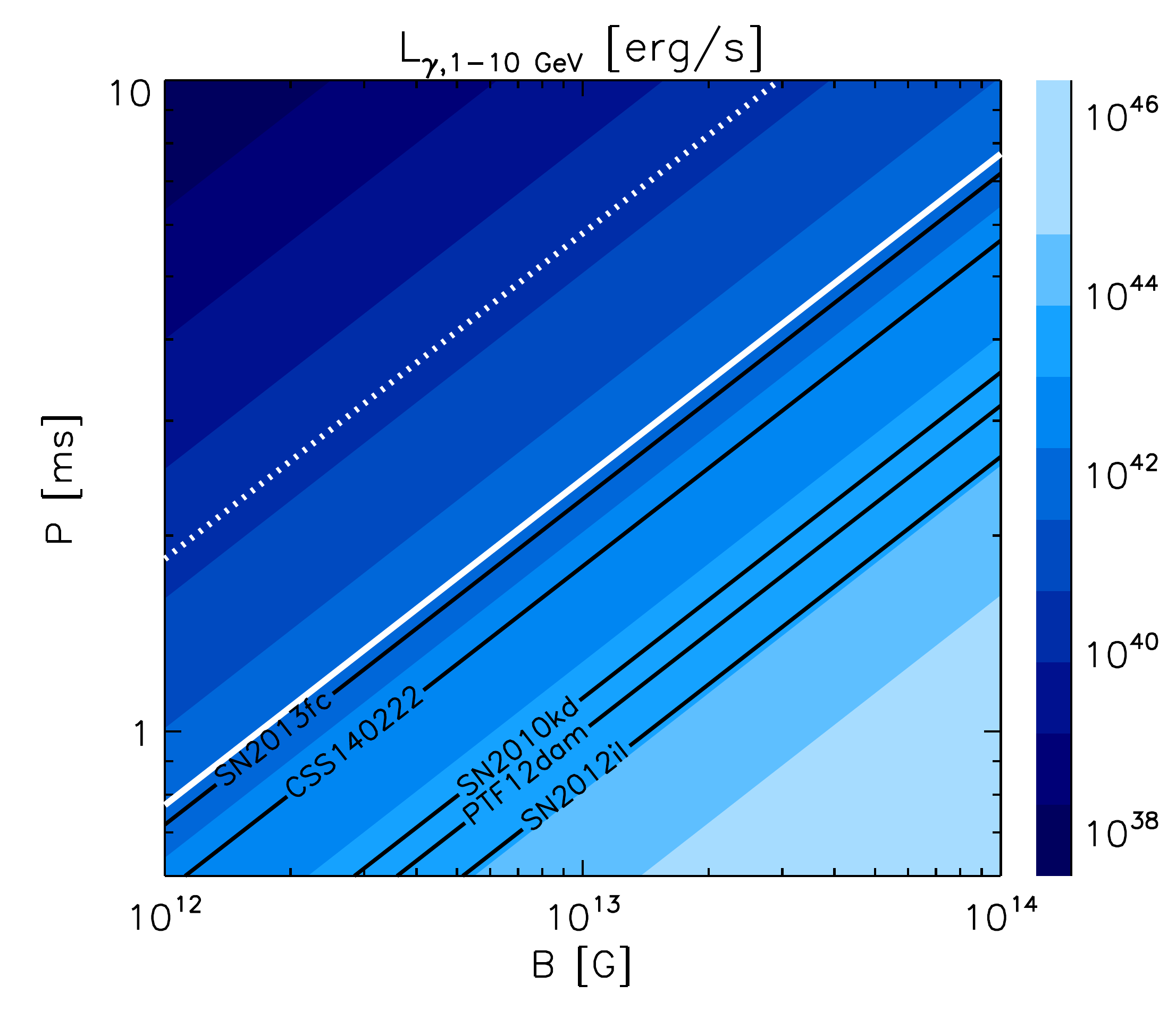}
\caption{Expected $\gamma$-ray luminosities $L_{\gamma,1-10\,{\rm GeV}}$ of SLSNe powered by neutron stars with dipole magnetic field $B$ and rotation period $P$, assuming a pair luminosity fraction $\eta_e=1$ and an attenuation factor $\xi=0.1$ (see Eq.~\ref{eq:Lgamma}). Overlaid are the $\gamma$-ray luminosity limits derived in this work for selected individual SLSNe in black lines, and for the total joint likelihood analysis sample as a white line (Table~\ref{tab_SNe_fluxes_stacking_all_src_7bnd}), for time windows of $t_{\rm SN}$ to $t_{\rm SN}+2\,$years or $t_{\rm SN}+1\,$year for \object{CSS140222}. The white dotted line indicates the limit derived for standard core-collapse SNe \citep{Fermi_CSM15}. The parameter-space above the line is allowed for a given source or population, modulo the scaling factor $\eta_e\,\xi_{-1}$, with $\xi <1$ in any case. The high magnetic field end ($B\gtrsim 5\times10^{13}\,$G) should be viewed with care as in the magnetar regime, $\xi\ll 1$, implying less stringent constraints on the parameter space.}
\label{fig_Lumcont}
\end{figure}

From the upper limits on the $\gamma$-ray luminosity measured at the location of SLSNe, it is possible to derive constraints on the values of the central neutron star period $P$ and dipole field strength $B$. 

The emitted $\gamma$-ray signal is produced by leptons accelerated in the young neutron star wind nebula region, as is evidenced for example in the Crab nebula and as is successfully modelled in various other pulsar wind nebulae (e.g. \citealp{Gelfand09,Fang10,Bucciantini11,Tanaka11}). The energy channeled into \gam-rays should scale as the electromagnetic luminosity provided by the pulsar into the wind\footnote{In this section, numerical quantities are noted $Q_x\equiv Q/10^x$ in cgs units, unless specified otherwise.
}, $L_{\rm p}= {L_{\rm rot}}/{(1+t/t_{\rm  p})^2}$, where the initial pulsar luminosity $L_{\rm rot}=E_{\rm  rot}/t_{\rm p}\,\sim\,0.64\times10^{45} P_{-3}^{-4}B_{13}^2R_{6}^6\;$erg/s, over a typical spin-down timescale $t_{\rm p}$. The pulsar rotational energy reservoir can be written $E_{\rm rot} ={2\pi^2IP^{-2}} \sim 2.0\times 10^{5f2}\;{\rm erg}\, I_{45} P_{\rm i,-3}^{-2}$, assuming for simplicity a pulsar spin-down braking index\footnote{Although observations indicate $n\,\sim2-2.5$, our choice of breaking index does not impact our results, as most of the neutron star rotational energy has been released at times $t_{\rm p}$, at which we make our measurements.} $n=3$. The spin-down timescale is given by $t_{\rm p} \sim
3.1\times 10^{7}\,{\rm  s}\,I_{45}B_{13}^{-2}R_{6}^{-6}P_{-3}^2$. In all the above formulae, $R$ and $I$ are the star radius and the moment of inertia (see \citealp{Shapiro83}), respectively.

A fraction $\eta_e$ of $L_p$ is dissipated into pairs at the pulsar wind nebula. The level of this dissipation is currently the subject of intense discussions in the community, and is related to the so-called \textquotedblleft sigma-problem\textquotedblright ~(see e.g. Kirk et al. 2009). However, the observations of the Crab nebula and of other young nebulae point to $\eta_e\sim 1$ (e.g. \citealp{Kirk09}) with a less significant fraction of the wind energy going into the nebula magnetic field. 
The pairs then radiate via synchrotron and inverse Compton (IC) processes in the nebula region, and this emission is attenuated by the radiation fields in the nebula and by matter further out in the supernova ejecta. 

In the $1-10\,$GeV energy range, for observation times of months to a few years after the supernova explosion, the radiation is dominated by the IC process and the obtained spectrum follows approximately a power law of index $\sim -2$ \citep{Murase14}. The expected luminosity of a young neutron star at time $t_{\rm p}$ at  energies $\epsilon\sim 1-10\,$GeV can then be written \citep{KPO13,Murase14} as 
\begin{eqnarray}\label{eq:Lgamma}
L_{\gamma,\epsilon}&\sim &\xi \eta_e Y(1+Y)^{-1}\,{L_{\rm rot}} \\
&\sim& 3.2\times10^{44} \,\eta_e\,\xi_{-1}\,P_{-3}^{-4}B_{13}^2R_{6}^6\;{\rm erg/s}\ .\nonumber
\end{eqnarray}
Here $Y=t_{\rm syn}/t_{\rm IC}$ is the Compton parameter, i.e. the ratio between the synchrotron and IC cooling timescales. 
The numerical value is calculated assuming a Thomson regime $Y=1$. This is a safe estimate as the Klein-Nishina effect mainly cuts off the flux at high energies but does not affect the overall normalization between $1-10\,$GeV. 
The factor $\xi<1$ takes into account the spread of the IC radiation over a given energy range with the uncertainties on the spectral indices at injection---which can range from hard indices $\sim -1.5$ for reconnection-type, one-shot acceleration processes to softer indices $\lesssim -2$ for stochastic acceleration mechanisms---and the attenuation due to radiation and matter. Interestingly, in this energy range, for $t\sim t_{\rm p}$ and for mildly magnetized objects ($B\gtrsim 5\times10^{13}\,$G), the radiated flux is robust to attenuation by the nebula radiation fields and matter, within a factor of a few \citep{Murase15}. A value of $\xi=0.1$ can thus be viewed as reasonable. However, for magnetars, the radiated spectra are softer and the overall flux are lower and lead to $\xi \ll 1$. 

Figure~\ref{fig_Lumcont} shows the contours of the  luminosity $L_{\gamma,1-10\,{\rm GeV}}$ estimated in the parameter space $P-B$, assuming $\eta_e=1$ and $\xi=0.1$. We overlaid the luminosity limits derived for individual SLSNe for which we obtained the strongest constraints, and for the total stacked sample for time windows of $t_{\rm SN}$ to $t_{\rm SN}+2\,$years or $t_{\rm SN}+1\,$year. Table~\ref{tab_SNe_fluxes_7bnd_plot} summarizes these upper limits. In the contour plot, the sets of $P-B$ above the line are allowed for a given source or population, modulo the scaling factor $\eta_e\,\xi_{-1}$. In particular within the conservative hypotheses on $\eta_e$ and $\xi_{-1}$, the central pulsar can be sub-millisecond only if $B<2\times10^{12}$\,G. In Figs.~7 to 10 of \citet{Murase15}, two simulated cases for $P=2\,$ms and $P=10\,$ms are presented for different magnetic field values. The results presented in this paper rule out the case with $P=2\,$ms.

\begin{table}%[!h]
\centering 
\caption{Luminosity upper limits used in Fig. \ref{fig_Lumcont}. The second column contains the limits obtained on the sum of derived luminosities for the dataset duration indicated in the third column in the individual energy bands between 600 MeV and 10 GeV.} 
\label{tab_SNe_fluxes_7bnd_plot} 
%\resizebox{\textwidth}{!}{
\begin{tabular}{lcccccccc} 
Name & $L_{0.6-10.2\,\rm{GeV}}$ & Dataset duration \\ 
\noalign{\smallskip} 
 & [erg\,s$^{-1}$] & yr \\ 
\hline 
\noalign{\smallskip} 
\object{SN2010kd} & $<2.0\times10^{43}$ & 2 \\ 
\noalign{\smallskip} 
\object{SN2012il} & $<6.6\times10^{43}$ & 2 \\ 
\noalign{\smallskip} 
\object{PTF12dam} & $<3.2\times10^{43}$ & 2 \\ 
\noalign{\smallskip} 
\object{SN2013fc} & $<1.2\times10^{42}$ & 2 \\ 
\noalign{\smallskip} 
\object{CSS140222} & $<3.1\times10^{42}$ & 1 \\
\noalign{\smallskip} 
Full sample & $<9.1\times10^{41}$ & 2 \\ 
\noalign{\smallskip} 
\hline 
\end{tabular}%}
\end{table} 

The limit given by the joint likelihood analysis of all sources in our sample (white line) places strong constraints on the rate of the neutron star population with mild dipole magnetic fields and millisecond rotation periods. It indicates that the rate of objects born with millisecond-rotation periods $P\lesssim 2\,$ms and $B\sim10^{12-13}\,$G (where the assumption $\eta_e\,\xi_{-1}=1$ is conservative) must be lower than the rate of the observed SLSNe (of order $\sim 200\,$Gpc$^{-3}\,$yr$^{-1}$ at $z\sim 0.16$; \citealp{Quimby13}). The luminosity limits obtained on individual sources are also constraining: in particular, \object{SN2013fc}, \object{CSS140222}, and \object{SN2010kd} can be born with millisecond periods only for $B\lesssim10^{13}\,$G. 
The derived upper limit for \object{SN2013fc}, the closest source of the sample located at about 80Mpc, is only $\sim$30\% higher than for the joint likelihood analysis (whose reference distance is equal to 100\,Mpc) for the same time window. This indicates that the combined limit, obtained with the $1/d^2$-weighting, is dominated by the closest source(s). A similar result was noticed in \cite{Fermi_CSM15}.

\cite{Fermi_CSM15} followed the same method of joint likelihood analysis to search for the emission from standard core-collapse supernovae. They discuss their results within the framework of the model of interaction of the SN ejecta with the circumstellar material. As in this work, weighting for the distances, they derived upper limits on the emitted luminosity. These authors obtain $L_{\gamma,1-10\,{\rm GeV}}< 2.8\times10^{40}\,$erg\,s$^{-1}$ compared to $L_{\gamma,1-10\,{\rm GeV}}<9.1\times10^{41}\,$erg\,s$^{-1}$ in our study. Their luminosity constraint is roughly a factor 30 tighter than that measured in this paper, despite a dataset that is a third shorter because they studied a sample more than three times larger and roughly five times closer (hence more detectable).

Our results suppose that the dipole magnetic field of the neutron star is set at its highest value at birth. However some studies \citep{1995ApJ...440L..77M,2011MNRAS.414.2567H,2012MNRAS.425.2487V} propose that the fallback accretion after a supernova explosion onto the newborn neutron star would result in the burial of the magnetic field into the crust and its re-emergence over a time scale of thousands years or more \citep[e.g.][]{1999A&A...345..847G, 2011MNRAS.414.2567H}. The diffusion time of the magnetic field, which results in its growth, is strongly dependent on the depth of burial, itself directly related to the mass of accreted matter \citep{2011MNRAS.414.2567H,1993PhRvL..70..379L}. In addition, the mass of accreted matter inversely scales with the space velocity of the neutron star \citep{2013MNRAS.430L..59G}. Hence according to this model, a runaway neutron star would be more unlikely to have a buried magnetic field. \cite{2016MNRAS.456.3813T} showed that masses as low as $10^{-3}-10^{-2}\,$M$_{\odot}$ are sufficient to bury a few $10^{12}\,$G magnetic field. This makes the occurrence of such phenomena not unusual and hence must be kept in mind when considering the pulsar-aided scenario.
%%%%%%%%%%%%%%%%%%%%%%%%%%%%%%%%%%%%%%%%%%%%%%%%%%%
%--------------------------------------------------------%
%%%%%%%%%%%%%%%%%%%%%%%%%%%%%%%%%%%%%%%%%%%%%%%%%%%}
\section{Conclusion}
We searched for the first time, through individual and stacking analyses, for $\gamma$-ray emission from a reasonable sample of SLSNe discovered through optical surveys. No signals were observed above the detection threshold and we derived the first upper limits on $\gamma$-ray signals from these objects. Assuming a scaling of the $\gamma$-ray flux with $1/d^2$, we report an upper limit at 95\% CL to the $\gamma$-ray luminosity $L_{\gamma}<9.1\times10^{41}$\,erg\,s$^{-1}$ for an assumed $E^{-2}$ photon spectrum, for our full SLSN sample and the two-year time window. 

Three scenarios are mainly proposed to explain the exceptional luminosities of SLSNe. Two of these scenarios, one relying on the interaction of the supernova ejecta with the circumstellar material and the other on the power supplied by a central compact object, predict $\gamma$-ray emission in the GeV-TeV range. Both can apply to SLSNe but also to standard core-collapse SNe. 

From the LAT non-detection and the predictions from the neutron-star powered model, one can obtain observational constraints on the rotation period and dipolar magnetic field strength of the central object.
Based on conservative assumptions,  
we find that the rate of the neutron stars born with millisecond rotation periods {$P\lesssim 2\,$ms and $B\sim10^{12-13}\,$G} must be lower than the rate of the observed SLSNe. The luminosity limits obtained on some individual sources are also constraining. 

We recommend reiterating this analysis in the future with a much larger sample of SLSNe (upper limits decreasing as the square root of the number of stacked sources), more $\gamma$-ray data and better sensitivity. However it would be really difficult to improve the upper limits by more than a factor 2 or 3. Another approach would be to weight the sources differently, with respect to the optical flux for instance as in \cite{Fermi_CSM15}, if one is able to consistently concatenate a catalogue of optical flux values for a large sample of SLSNe. In any case, future studies of SLSNe will benefit from the upcoming optical surveys that will provide an unprecedently complete catalogue of detailed informations on SNe. The Zwicky Transient Facility \citep[ZTF; first light in 2017,][]{ZTF} and the Large Synoptic Survey Telescope \citep[LSST; under construction in Chile,][]{LSST} will be particularly relevant and efficient.\\
%%%%%%%%%%%%%%%%%%%%%%%%%%%%%%%%%%%%%%%%%%%%%%%%%%%
%--------------------------------------------------------%
%%%%%%%%%%%%%%%%%%%%%%%%%%%%%%%%%%%%%%%%%%%%%%%%%%%
\begin{acknowledgements}
We thank K. Murase for fruitful discussions and A. Franckowiak for her careful reading and insightful comments. NRT was supported by the PER-SU fellowship at Sorbonne Universit\'es. 
KK acknowledges financial support from the PER-SU fellowship at Sorbonne Universit\'es and from the Labex ILP (reference ANR-10-LABX-63, ANR-11-IDEX-0004-02). 
This work is supported by the APACHE grant (ANR-16-CE31-0001) of the French Agence Nationale de la Recherche.
The \textit{Fermi}-LAT Collaboration acknowledges generous ongoing support
from a number of agencies and institutes that have supported both the
development and operation of the LAT as well as scientific data analysis.
These include the National Aeronautics and Space Administration and the
Department of Energy in the United States, the Commissariat \`a l'Energie Atomique
and the Centre National de la Recherche Scientifique / Institut National de Physique
Nucl\'eaire et de Physique des Particules in France, the Agenzia Spaziale Italiana
and the Istituto Nazionale di Fisica Nucleare in Italy, the Ministry of Education,
Culture, Sports, Science and Technology (MEXT), High Energy Accelerator Research
Organization (KEK) and Japan Aerospace Exploration Agency (JAXA) in Japan, and
the K.~A.~Wallenberg Foundation, the Swedish Research Council and the
Swedish National Space Board in Sweden.
Additional support for science analysis during the operations phase is gratefully acknowledged from the Istituto Nazionale di Astrofisica in Italy and the Centre National d'\'Etudes Spatiales in France.
S. A. was supported by Netherlands Organization for Scientific Research (NWO) through a Vidi grant. This work performed in part under DOE Contract DE-AC02-76SF00515.
We also thank \cite{2017ApJ...835...64G} for the useful online catalogue of SN data they set up.
The authors wish to acknowledge the anonymous referee for the helpful suggestions and comments that enriched the paper and helped to highlight some of its results.
\end{acknowledgements}

\bibliographystyle{aa}  
\bibliography{SLSNe}

\begin{thebibliography}{101}
\expandafter\ifx\csname natexlab\endcsname\relax\def\natexlab#1{#1}\fi

\bibitem[{{Abell} {et~al.}(2009){Abell}, {Allison}, {Anderson}, {Andrew},
  {Angel}, {Armus}, {Arnett}, {Asztalos}, {Axelrod}, \& et~al.}]{LSST}
{Abell}, P.~A., {Allison}, J., {Anderson}, S.~F., {et~al.} 2009, ArXiv
  e-prints, LSST Science Book, Version 2.0 [\eprint[arXiv]{0912.0201}]

\bibitem[{{Acero} {et~al.}(2015){Acero}, {Ackermann}, {Ajello}, {Albert},
  {Atwood}, {Axelsson}, {Baldini}, {Ballet}, {Barbiellini}, {Bastieri},
  {Belfiore}, {Bellazzini}, {Bissaldi}, {Blandford}, {Bloom}, {Bogart},
  {Bonino}, {Bottacini}, {Bregeon}, {Britto}, {Bruel}, {Buehler}, {Burnett},
  {Buson}, {Caliandro}, {Cameron}, {Caputo}, {Caragiulo}, {Caraveo},
  {Casandjian}, {Cavazzuti}, {Charles}, {Chaves}, {Chekhtman}, {Cheung},
  {Chiang}, {Chiaro}, {Ciprini}, {Claus}, {Cohen-Tanugi}, {Cominsky}, {Conrad},
  {Cutini}, {D'Ammando}, {de Angelis}, {DeKlotz}, {de Palma}, {Desiante},
  {Digel}, {Di Venere}, {Drell}, {Dubois}, {Dumora}, {Favuzzi}, {Fegan},
  {Ferrara}, {Finke}, {Franckowiak}, {Fukazawa}, {Funk}, {Fusco}, {Gargano},
  {Gasparrini}, {Giebels}, {Giglietto}, {Giommi}, {Giordano}, {Giroletti},
  {Glanzman}, {Godfrey}, {Grenier}, {Grondin}, {Grove}, {Guillemot}, {Guiriec},
  {Hadasch}, {Harding}, {Hays}, {Hewitt}, {Hill}, {Horan}, {Iafrate}, {Jogler},
  {J{\'o}hannesson}, {Johnson}, {Johnson}, {Johnson}, {Johnson}, {Kamae},
  {Kataoka}, {Katsuta}, {Kuss}, {La Mura}, {Landriu}, {Larsson}, {Latronico},
  {Lemoine-Goumard}, {Li}, {Li}, {Longo}, {Loparco}, {Lott}, {Lovellette},
  {Lubrano}, {Madejski}, {Massaro}, {Mayer}, {Mazziotta}, {McEnery},
  {Michelson}, {Mirabal}, {Mizuno}, {Moiseev}, {Mongelli}, {Monzani},
  {Morselli}, {Moskalenko}, {Murgia}, {Nuss}, {Ohno}, {Ohsugi}, {Omodei},
  {Orienti}, {Orlando}, {Ormes}, {Paneque}, {Panetta}, {Perkins},
  {Pesce-Rollins}, {Piron}, {Pivato}, {Porter}, {Racusin}, {Rando}, {Razzano},
  {Razzaque}, {Reimer}, {Reimer}, {Reposeur}, {Rochester}, {Romani},
  {Salvetti}, {S{\'a}nchez-Conde}, {Saz Parkinson}, {Schulz}, {Siskind},
  {Smith}, {Spada}, {Spandre}, {Spinelli}, {Stephens}, {Strong}, {Suson},
  {Takahashi}, {Takahashi}, {Tanaka}, {Thayer}, {Thayer}, {Thompson},
  {Tibaldo}, {Tibolla}, {Torres}, {Torresi}, {Tosti}, {Troja}, {Van Klaveren},
  {Vianello}, {Winer}, {Wood}, {Wood}, \& {Zimmer}}]{3FGL}
{Acero}, F., {Ackermann}, M., {Ajello}, M., {et~al.} 2015, \apjs, 218, 23

\bibitem[{{Acero} {et~al.}(2016{\natexlab{a}}){Acero}, {Ackermann}, {Ajello},
  {Albert}, {Baldini}, {Ballet}, {Barbiellini}, {Bastieri}, {Bellazzini},
  {Bissaldi}, {Bloom}, {Bonino}, {Bottacini}, {Brandt}, {Bregeon}, {Bruel},
  {Buehler}, {Buson}, {Caliandro}, {Cameron}, {Caragiulo}, {Caraveo},
  {Casandjian}, {Cavazzuti}, {Cecchi}, {Charles}, {Chekhtman}, {Chiang},
  {Chiaro}, {Ciprini}, {Claus}, {Cohen-Tanugi}, {Conrad}, {Cuoco}, {Cutini},
  {D'Ammando}, {de Angelis}, {de Palma}, {Desiante}, {Digel}, {Di Venere},
  {Drell}, {Favuzzi}, {Fegan}, {Ferrara}, {Focke}, {Franckowiak}, {Funk},
  {Fusco}, {Gargano}, {Gasparrini}, {Giglietto}, {Giordano}, {Giroletti},
  {Glanzman}, {Godfrey}, {Grenier}, {Guiriec}, {Hadasch}, {Harding}, {Hayashi},
  {Hays}, {Hewitt}, {Hill}, {Horan}, {Hou}, {Jogler}, {J{\'o}hannesson},
  {Kamae}, {Kuss}, {Landriu}, {Larsson}, {Latronico}, {Li}, {Li}, {Longo},
  {Loparco}, {Lovellette}, {Lubrano}, {Maldera}, {Malyshev}, {Manfreda},
  {Martin}, {Mayer}, {Mazziotta}, {McEnery}, {Michelson}, {Mirabal}, {Mizuno},
  {Monzani}, {Morselli}, {Nuss}, {Ohsugi}, {Omodei}, {Orienti}, {Orlando},
  {Ormes}, {Paneque}, {Pesce-Rollins}, {Piron}, {Pivato}, {Rain{\`o}}, {Rando},
  {Razzano}, {Razzaque}, {Reimer}, {Reimer}, {Remy}, {Renault},
  {S{\'a}nchez-Conde}, {Schaal}, {Schulz}, {Sgr{\`o}}, {Siskind}, {Spada},
  {Spandre}, {Spinelli}, {Strong}, {Suson}, {Tajima}, {Takahashi}, {Thayer},
  {Thompson}, {Tibaldo}, {Tinivella}, {Torres}, {Tosti}, {Troja}, {Vianello},
  {Werner}, {Wood}, {Wood}, {Zaharijas}, \& {Zimmer}}]{2016ApJS..223...26A}
{Acero}, F., {Ackermann}, M., {Ajello}, M., {et~al.} 2016{\natexlab{a}}, \apjs,
  223, 26

\bibitem[{{Acero} {et~al.}(2016{\natexlab{b}}){Acero}, {Ackermann}, {Ajello},
  {Baldini}, {Ballet}, {Barbiellini}, {Bastieri}, {Bellazzini}, {Bissaldi},
  {Blandford}, {Bloom}, {Bonino}, {Bottacini}, {Brandt}, {Bregeon}, {Bruel},
  {Buehler}, {Buson}, {Caliandro}, {Cameron}, {Caputo}, {Caragiulo}, {Caraveo},
  {Casandjian}, {Cavazzuti}, {Cecchi}, {Chekhtman}, {Chiang}, {Chiaro},
  {Ciprini}, {Claus}, {Cohen}, {Cohen-Tanugi}, {Cominsky}, {Condon}, {Conrad},
  {Cutini}, {D'Ammando}, {de Angelis}, {de Palma}, {Desiante}, {Digel}, {Di
  Venere}, {Drell}, {Drlica-Wagner}, {Favuzzi}, {Ferrara}, {Franckowiak},
  {Fukazawa}, {Funk}, {Fusco}, {Gargano}, {Gasparrini}, {Giglietto}, {Giommi},
  {Giordano}, {Giroletti}, {Glanzman}, {Godfrey}, {Gomez-Vargas}, {Grenier},
  {Grondin}, {Guillemot}, {Guiriec}, {Gustafsson}, {Hadasch}, {Harding},
  {Hayashida}, {Hays}, {Hewitt}, {Hill}, {Horan}, {Hou}, {Iafrate}, {Jogler},
  {J{\'o}hannesson}, {Johnson}, {Kamae}, {Katagiri}, {Kataoka}, {Katsuta},
  {Kerr}, {Kn{\"o}dlseder}, {Kocevski}, {Kuss}, {Laffon}, {Lande}, {Larsson},
  {Latronico}, {Lemoine-Goumard}, {Li}, {Li}, {Longo}, {Loparco}, {Lovellette},
  {Lubrano}, {Magill}, {Maldera}, {Marelli}, {Mayer}, {Mazziotta}, {Michelson},
  {Mitthumsiri}, {Mizuno}, {Moiseev}, {Monzani}, {Moretti}, {Morselli},
  {Moskalenko}, {Murgia}, {Nemmen}, {Nuss}, {Ohsugi}, {Omodei}, {Orienti},
  {Orlando}, {Ormes}, {Paneque}, {Perkins}, {Pesce-Rollins}, {Petrosian},
  {Piron}, {Pivato}, {Porter}, {Rain{\`o}}, {Rando}, {Razzano}, {Razzaque},
  {Reimer}, {Reimer}, {Renaud}, {Reposeur}, {Rousseau}, {Saz Parkinson},
  {Schmid}, {Schulz}, {Sgr{\`o}}, {Siskind}, {Spada}, {Spandre}, {Spinelli},
  {Strong}, {Suson}, {Tajima}, {Takahashi}, {Tanaka}, {Thayer}, {Thompson},
  {Tibaldo}, {Tibolla}, {Torres}, {Tosti}, {Troja}, {Uchiyama}, {Vianello},
  {Wells}, {Wood}, {Wood}, {Yassine}, {den Hartog}, \&
  {Zimmer}}]{1st_Fermi_SNR_cat}
{Acero}, F., {Ackermann}, M., {Ajello}, M., {et~al.} 2016{\natexlab{b}}, \apjs,
  224, 8

\bibitem[{{Ackermann} {et~al.}(2015){Ackermann}, {Arcavi}, {Baldini}, {Ballet},
  {Barbiellini}, {Bastieri}, {Bellazzini}, {Bissaldi}, {Blandford}, {Bonino},
  {Bottacini}, {Brandt}, {Bregeon}, {Bruel}, {Buehler}, {Buson}, {Caliandro},
  {Cameron}, {Caragiulo}, {Caraveo}, {Cavazzuti}, {Cecchi}, {Charles},
  {Chekhtman}, {Chiang}, {Chiaro}, {Ciprini}, {Claus}, {Cohen-Tanugi},
  {Cutini}, {D'Ammando}, {de Angelis}, {de Palma}, {Desiante}, {Di Venere},
  {Drell}, {Favuzzi}, {Fegan}, {Franckowiak}, {Funk}, {Fusco}, {Gal-Yam},
  {Gargano}, {Gasparrini}, {Giglietto}, {Giordano}, {Giroletti}, {Glanzman},
  {Godfrey}, {Grenier}, {Grove}, {Guiriec}, {Harding}, {Hayashi}, {Hewitt},
  {Hill}, {Horan}, {Jogler}, {J{\'o}hannesson}, {Kocevski}, {Kuss}, {Larsson},
  {Lashner}, {Latronico}, {Li}, {Li}, {Longo}, {Loparco}, {Lovellette},
  {Lubrano}, {Malyshev}, {Mayer}, {Mazziotta}, {McEnery}, {Michelson},
  {Mizuno}, {Monzani}, {Morselli}, {Murase}, {Nugent}, {Nuss}, {Ofek},
  {Ohsugi}, {Orienti}, {Orlando}, {Ormes}, {Paneque}, {Pesce-Rollins}, {Piron},
  {Pivato}, {Rain{\`o}}, {Rando}, {Razzano}, {Reimer}, {Reimer}, {Schulz},
  {Sgr{\`o}}, {Siskind}, {Spada}, {Spandre}, {Spinelli}, {Suson}, {Takahashi},
  {Thayer}, {Tibaldo}, {Torres}, {Troja}, {Vianello}, {Werner}, {Wood}, \&
  {Wood}}]{Fermi_CSM15}
{Ackermann}, M., {Arcavi}, I., {Baldini}, L., {et~al.} 2015, \apj, 807, 169

\bibitem[{{Ade} {et~al.}(2015){Ade}, {Aghanim}, {Aniano}, {Arnaud}, {Ashdown},
  {Aumont}, {Baccigalupi}, {Banday}, {Barreiro}, {Bartolo}, \&
  et~al.}]{2015A&A...582A..31P}
{Ade}, P.~A.~R., {Aghanim}, N., {Aniano}, G., {et~al.} 2015, \aap, 582, A31

\bibitem[{{Ade} {et~al.}(2014){Ade}, {Aghanim}, {Armitage-Caplan}, {Arnaud},
  {Ashdown}, {Atrio-Barandela}, {Aumont}, {Baccigalupi}, {Banday}, \&
  et~al.}]{Planck2013_cosmoparam}
{Ade}, P.~A.~R., {Aghanim}, N., {Armitage-Caplan}, C., {et~al.} 2014, \aap,
  571, A16

\bibitem[{{Ade} {et~al.}(2016){Ade}, {Aghanim}, {Arnaud}, {Ashdown}, {Aumont},
  {Baccigalupi}, {Banday}, {Barreiro}, {Bartlett}, \&
  et~al.}]{Planck2015_cosmoparam}
{Ade}, P.~A.~R., {Aghanim}, N., {Arnaud}, M., {et~al.} 2016, \aap, 594, A13

\bibitem[{{Anderson} {et~al.}(2015){Anderson}, {Chiang}, {Cohen-Tanugi},
  {Conrad}, {Drlica-Wagner}, {Llena Garde}, \& {Stephan Zimmer for the Fermi
  Lat Collaboration}}]{2015arXiv150203081A}
{Anderson}, B., {Chiang}, J., {Cohen-Tanugi}, J., {et~al.} 2015, in 5th Fermi
  Symposium, 5th Fermi Symposium

\bibitem[{{Atwood} {et~al.}(2013){Atwood}, {Albert}, {Baldini}, {Tinivella},
  {Bregeon}, {Pesce-Rollins}, {Sgr{\`o}}, {Bruel}, {Charles}, {Drlica-Wagner},
  {Franckowiak}, {Jogler}, {Rochester}, {Usher}, {Wood}, {Cohen-Tanugi}, \&
  {S.~Zimmer for the Fermi-LAT Collaboration}}]{2013arXiv1303.3514A}
{Atwood}, W., {Albert}, A., {Baldini}, L., {et~al.} 2013, in 2012 Fermi
  Symposium proceedings

\bibitem[{{Atwood} {et~al.}(2009){Atwood}, {Abdo}, {Ackermann}, {Althouse},
  {Anderson}, {Axelsson}, {Baldini}, {Ballet}, {Band}, {Barbiellini},
  {Bartelt}, {Bastieri}, {Baughman}, {Bechtol}, {B{\'e}d{\'e}r{\`e}de},
  {Bellardi}, {Bellazzini}, {Berenji}, {Bignami}, {Bisello}, {Bissaldi},
  {Blandford}, {Bloom}, {Bogart}, {Bonamente}, {Bonnell}, {Borgland},
  {Bouvier}, {Bregeon}, {Brez}, {Brigida}, {Bruel}, {Burnett}, {Busetto},
  {Caliandro}, {Cameron}, {Caraveo}, {Carius}, {Carlson}, {Casandjian},
  {Cavazzuti}, {Ceccanti}, {Cecchi}, {Charles}, {Chekhtman}, {Cheung},
  {Chiang}, {Chipaux}, {Cillis}, {Ciprini}, {Claus}, {Cohen-Tanugi},
  {Condamoor}, {Conrad}, {Corbet}, {Corucci}, {Costamante}, {Cutini}, {Davis},
  {Decotigny}, {DeKlotz}, {Dermer}, {de Angelis}, {Digel}, {do Couto e Silva},
  {Drell}, {Dubois}, {Dumora}, {Edmonds}, {Fabiani}, {Farnier}, {Favuzzi},
  {Flath}, {Fleury}, {Focke}, {Funk}, {Fusco}, {Gargano}, {Gasparrini},
  {Gehrels}, {Gentit}, {Germani}, {Giebels}, {Giglietto}, {Giommi}, {Giordano},
  {Glanzman}, {Godfrey}, {Grenier}, {Grondin}, {Grove}, {Guillemot}, {Guiriec},
  {Haller}, {Harding}, {Hart}, {Hays}, {Healey}, {Hirayama}, {Hjalmarsdotter},
  {Horn}, {Hughes}, {J{\'o}hannesson}, {Johansson}, {Johnson}, {Johnson},
  {Johnson}, {Johnson}, {Kamae}, {Katagiri}, {Kataoka}, {Kavelaars}, {Kawai},
  {Kelly}, {Kerr}, {Klamra}, {Kn{\"o}dlseder}, {Kocian}, {Komin}, {Kuehn},
  {Kuss}, {Landriu}, {Latronico}, {Lee}, {Lee}, {Lemoine-Goumard}, {Lionetto},
  {Longo}, {Loparco}, {Lott}, {Lovellette}, {Lubrano}, {Madejski}, {Makeev},
  {Marangelli}, {Massai}, {Mazziotta}, {McEnery}, {Menon}, {Meurer},
  {Michelson}, {Minuti}, {Mirizzi}, {Mitthumsiri}, {Mizuno}, {Moiseev},
  {Monte}, {Monzani}, {Moretti}, {Morselli}, {Moskalenko}, {Murgia},
  {Nakamori}, {Nishino}, {Nolan}, {Norris}, {Nuss}, {Ohno}, {Ohsugi}, {Omodei},
  {Orlando}, {Ormes}, {Paccagnella}, {Paneque}, {Panetta}, {Parent}, {Pearce},
  {Pepe}, {Perazzo}, {Pesce-Rollins}, {Picozza}, {Pieri}, {Pinchera}, {Piron},
  {Porter}, {Poupard}, {Rain{\`o}}, {Rando}, {Rapposelli}, {Razzano}, {Reimer},
  {Reimer}, {Reposeur}, {Reyes}, {Ritz}, {Rochester}, {Rodriguez}, {Romani},
  {Roth}, {Russell}, {Ryde}, {Sabatini}, {Sadrozinski}, {Sanchez}, {Sander},
  {Sapozhnikov}, {Parkinson}, {Scargle}, {Schalk}, {Scolieri}, {Sgr{\`o}},
  {Share}, {Shaw}, {Shimokawabe}, {Shrader}, {Sierpowska-Bartosik}, {Siskind},
  {Smith}, {Smith}, {Spandre}, {Spinelli}, {Starck}, {Stephens}, {Strickman},
  {Strong}, {Suson}, {Tajima}, {Takahashi}, {Takahashi}, {Tanaka}, {Tenze},
  {Tether}, {Thayer}, {Thayer}, {Thompson}, {Tibaldo}, {Tibolla}, {Torres},
  {Tosti}, {Tramacere}, {Turri}, {Usher}, {Vilchez}, {Vitale}, {Wang},
  {Watters}, {Winer}, {Wood}, {Ylinen}, \& {Ziegler}}]{LATinstrument}
{Atwood}, W.~B., {Abdo}, A.~A., {Ackermann}, M., {et~al.} 2009, \apj, 697,
  1071, (LAT Instrument Paper)

\bibitem[{{Baltay} {et~al.}(2013){Baltay}, {Rabinowitz}, {Hadjiyska}, {Walker},
  {Nugent}, {Coppi}, {Ellman}, {Feindt}, {McKinnon}, {Horowitz}, \&
  {Effron}}]{2013PASP..125..683B}
{Baltay}, C., {Rabinowitz}, D., {Hadjiyska}, E., {et~al.} 2013, \pasp, 125, 683

\bibitem[{{Bellm}(2014)}]{ZTF}
{Bellm}, E. 2014, in The Third Hot-wiring the Transient Universe Workshop, ed.
  P.~R. {Wozniak}, M.~J. {Graham}, A.~A. {Mahabal}, \& R.~{Seaman}, 27--33

\bibitem[{{Benetti} {et~al.}(2014){Benetti}, {Nicholl}, {Cappellaro},
  {Pastorello}, {Smartt}, {Elias-Rosa}, {Drake}, {Tomasella}, {Turatto},
  {Harutyunyan}, {Taubenberger}, {Hachinger}, {Morales-Garoffolo}, {Chen},
  {Djorgovski}, {Fraser}, {Gal-Yam}, {Inserra}, {Mazzali}, {Pumo}, {Sollerman},
  {Valenti}, {Young}, {Dennefeld}, {Le Guillou}, {Fleury}, \&
  {L{\'e}get}}]{Benetti14}
{Benetti}, S., {Nicholl}, M., {Cappellaro}, E., {et~al.} 2014, MNRAS, 441, 289

\bibitem[{{Benitez} {et~al.}(2014){Benitez}, {Polshaw}, {Inserra},
  {Botticella}, {Walker}, {Benetti}, {Pastorello}, {Smartt}, {Smith}, {Young},
  {Sullivan}, {Taubenberger}, {Valenti}, {Fraser}, {Manulis}, {Yaron},
  {Gal-Yam}, {Knapic}, {Smareglia}, {Molinaro}, {Baltay}, {Ellman},
  {Hadjiyska}, {McKinnon}, \& {Rabinowitz}}]{2014ATel.6118....1B}
{Benitez}, S., {Polshaw}, J., {Inserra}, C., {et~al.} 2014, The Astronomer's
  Telegram, 6118

\bibitem[{{Blagorodnova} {et~al.}(2014){Blagorodnova}, {Campbell}, {Fraser},
  {Walton}, {Anderson}, {Benetti}, {Pastorello}, {Inserra}, {Smartt}, {Smith},
  {Young}, {Sullivan}, {Taubenberger}, {Valenti}, {Yaron}, {Gal-Yam}, {Knapic},
  {Smareglia}, {Molinaro}, {Manulis}, {Baltay}, {Ellman}, {Hadjiyska},
  {McKinnon}, {Rabinowitz}, {Walker}, {Feindt}, {Kowalski}, {Nugent}, {Wright},
  {Kotak}, {Valenti}, {Burgett}, {Chambers}, {Huber}, {Kudritzki}, {Magnier},
  {Morgan}, {Stubbs}, {Sweeney}, {Tonry}, {Waters}, {Draper}, {Metcalfe},
  {Rest}, \& {Wyrzykowski}}]{2014ATel.5934....1B}
{Blagorodnova}, N., {Campbell}, H., {Fraser}, M., {et~al.} 2014, The
  Astronomer's Telegram, 5934

\bibitem[{{Bucciantini} {et~al.}(2011){Bucciantini}, {Arons}, \&
  {Amato}}]{Bucciantini11}
{Bucciantini}, N., {Arons}, J., \& {Amato}, E. 2011, MNRAS, 410, 381

\bibitem[{{Cano} {et~al.}(2015){Cano}, {de Ugarte Postigo}, {Perley},
  {Kr{\"u}hler}, {Margutti}, {Friis}, {Malesani}, {Jakobsson}, {Fynbo},
  {Gorosabel}, {Hjorth}, {S{\'a}nchez-Ram{\'{\i}}rez}, {Schulze}, {Tanvir},
  {Th{\"o}ne}, \& {Xu}}]{2015MNRAS.452.1535C}
{Cano}, Z., {de Ugarte Postigo}, A., {Perley}, D., {et~al.} 2015, \mnras, 452,
  1535

\bibitem[{{Cenko} {et~al.}(2010){Cenko}, {Kandrashoff}, {Silverman}, \&
  {Filippenko}}]{2010CBET.2461....2C}
{Cenko}, S.~B., {Kandrashoff}, M.~T., {Silverman}, J.~M., \& {Filippenko},
  A.~V. 2010, Central Bureau Electronic Telegrams, 2461, 2

\bibitem[{{Chandra} {et~al.}(2009){Chandra}, {Ofek}, {Frail}, {Quimby},
  {Kasliwal}, \& {Kulkarni}}]{2009ATel.2241....1C}
{Chandra}, P., {Ofek}, E.~O., {Frail}, D.~A., {et~al.} 2009, The Astronomer's
  Telegram, 2241

\bibitem[{{Chevalier} \& {Irwin}(2011)}]{Chevalier11}
{Chevalier}, R.~A. \& {Irwin}, C.~M. 2011, ApJ Lett., 729, L6

\bibitem[{{Chomiuk} {et~al.}(2011){Chomiuk}, {Chornock}, {Soderberg}, {Berger},
  {Chevalier}, {Foley}, {Huber}, {Narayan}, {Rest}, {Gezari}, {Kirshner},
  {Riess}, {Rodney}, {Smartt}, {Stubbs}, {Tonry}, {Wood-Vasey}, {Burgett},
  {Chambers}, {Czekala}, {Flewelling}, {Forster}, {Kaiser}, {Kudritzki},
  {Magnier}, {Martin}, {Morgan}, {Neill}, {Price}, {Roth}, {Sanders}, \&
  {Wainscoat}}]{2011ApJ...743..114C}
{Chomiuk}, L., {Chornock}, R., {Soderberg}, A.~M., {et~al.} 2011, \apj, 743,
  114

\bibitem[{{Choudalakis}(2011)}]{2011arXiv1101.0390C}
{Choudalakis}, G. 2011, prepared for PHYSTAT2011, ArXiv e-prints
  [\eprint[arXiv]{1101.0390}]

\bibitem[{{Cooke} {et~al.}(2012){Cooke}, {Sullivan}, {Gal-Yam}, {Barton},
  {Carlberg}, {Ryan-Weber}, {Horst}, {Omori}, \&
  {D{\'{\i}}az}}]{2012Natur.491..228C}
{Cooke}, J., {Sullivan}, M., {Gal-Yam}, A., {et~al.} 2012, \nat, 491, 228

\bibitem[{{Dessart} {et~al.}(2012){Dessart}, {Hillier}, {Waldman}, {Livne}, \&
  {Blondin}}]{Dessart12_2}
{Dessart}, L., {Hillier}, D.~J., {Waldman}, R., {Livne}, E., \& {Blondin}, S.
  2012, MNRAS, 426, L76

\bibitem[{{Dong} {et~al.}(2016){Dong}, {Shappee}, {Prieto}, {Jha}, {Stanek},
  {Holoien}, {Kochanek}, {Thompson}, {Morrell}, {Thompson}, {Basu}, {Beacom},
  {Bersier}, {Brimacombe}, {Brown}, {Bufano}, {Chen}, {Conseil}, {Danilet},
  {Falco}, {Grupe}, {Kiyota}, {Masi}, {Nicholls}, {Olivares E.}, {Pignata},
  {Pojmanski}, {Simonian}, {Szczygiel}, \& {Wo{\'z}niak}}]{2016Sci...351..257D}
{Dong}, S., {Shappee}, B.~J., {Prieto}, J.~L., {et~al.} 2016, Science, 351, 257

\bibitem[{{Drake} {et~al.}(2013{\natexlab{a}}){Drake}, {Djorgovski}, {Graham},
  {Mahabal}, {Williams}, {Prieto}, {Catelan}, {Larson}, {Christensen},
  {Inserra}, {Smartt}, {Fraser}, {Young}, {Smith}, {Wright}, {Kotak}, {McCrum},
  {Magill}, {Chen}, {Pastorello}, {Benetti}, {Valenti}, {Bresolin},
  {Kudritzki}, {Tonry}, {Magnier}, {Huber}, {Chambers}, {Kaiser}, {Morgan},
  {Burgett}, {Heasley}, {Sweeney}, {Waters}, {Flewelling}, {Stubbs}, {Price},
  {Sollerman}, {Taddia}, {Ergon}, {Leloudas}, \&
  {Taubenberger}}]{2013CBET.3459....1D}
{Drake}, A.~J., {Djorgovski}, S.~G., {Graham}, M.~J., {et~al.}
  2013{\natexlab{a}}, Central Bureau Electronic Telegrams, 3459

\bibitem[{{Drake} {et~al.}(2009{\natexlab{a}}){Drake}, {Djorgovski}, {Mahabal},
  {Beshore}, {Larson}, {Graham}, {Williams}, {Christensen}, {Catelan},
  {Boattini}, {Gibbs}, {Hill}, \& {Kowalski}}]{2009ApJ...696..870D}
{Drake}, A.~J., {Djorgovski}, S.~G., {Mahabal}, A., {et~al.}
  2009{\natexlab{a}}, \apj, 696, 870

\bibitem[{{Drake} {et~al.}(2009{\natexlab{b}}){Drake}, {Djorgovski}, {Mahabal},
  {Graham}, {Williams}, {Catelan}, {Beshore}, {Larson}, \&
  {Christensen}}]{2009CBET.1958....1D}
{Drake}, A.~J., {Djorgovski}, S.~G., {Mahabal}, A., {et~al.}
  2009{\natexlab{b}}, Central Bureau Electronic Telegrams, 1958

\bibitem[{{Drake} {et~al.}(2013{\natexlab{b}}){Drake}, {Djorgovski}, {Mahabal},
  {Williams}, {Prieto}, {Catelan}, {Christensen}, {Larson}, {Smartt},
  {Nicholl}, {Inserra}, {Wright}, \& {Chen}}]{2013CBET.3560....1D}
{Drake}, A.~J., {Djorgovski}, S.~G., {Mahabal}, A., {et~al.}
  2013{\natexlab{b}}, Central Bureau Electronic Telegrams, 3560

\bibitem[{{Drake} {et~al.}(2010{\natexlab{a}}){Drake}, {Djorgovski}, {Prieto},
  {Mahabal}, {Balam}, {Williams}, {Graham}, {Catelan}, {Beshore}, \&
  {Larson}}]{2010ApJ...718L.127D}
{Drake}, A.~J., {Djorgovski}, S.~G., {Prieto}, J.~L., {et~al.}
  2010{\natexlab{a}}, \apjl, 718, L127

\bibitem[{{Drake} {et~al.}(2010{\natexlab{b}}){Drake}, {Mahabal}, {Djorgovski},
  {Graham}, {Williams}, {Mohan}, {Ravindranath}, {Prieto}, {Ho}, {Kewley},
  {Myers}, {Catelan}, {Christensen}, {Beshore}, \&
  {Larson}}]{2010ATel.2544....1D}
{Drake}, A.~J., {Mahabal}, A.~A., {Djorgovski}, S.~G., {et~al.}
  2010{\natexlab{b}}, The Astronomer's Telegram, 2544

\bibitem[{{Fang} \& {Zhang}(2010)}]{Fang10}
{Fang}, J. \& {Zhang}, L. 2010, A\&A, 515, A20

\bibitem[{{Gal-Yam}(2012{\natexlab{a}})}]{Gal-Yam12}
{Gal-Yam}, A. 2012{\natexlab{a}}, Science, 337, 927

\bibitem[{{Gal-Yam}(2012{\natexlab{b}})}]{2012Sci...337..927G}
{Gal-Yam}, A. 2012{\natexlab{b}}, Science, 337, 927

\bibitem[{{Gal-Yam} \& {Leonard}(2009)}]{Gal-Yam09b}
{Gal-Yam}, A. \& {Leonard}, D.~C. 2009, Nature, 458, 865

\bibitem[{{Gal-Yam} {et~al.}(2009){Gal-Yam}, {Mazzali}, {Ofek}, {Nugent},
  {Kulkarni}, {Kasliwal}, {Quimby}, {Filippenko}, {Cenko}, {Chornock},
  {Waldman}, {Kasen}, {Sullivan}, {Beshore}, {Drake}, {Thomas}, {Bloom},
  {Poznanski}, {Miller}, {Foley}, {Silverman}, {Arcavi}, {Ellis}, \&
  {Deng}}]{Gal-Yam09}
{Gal-Yam}, A., {Mazzali}, P., {Ofek}, E.~O., {et~al.} 2009, Nature, 462, 624

\bibitem[{{Gelfand} {et~al.}(2009){Gelfand}, {Slane}, \& {Zhang}}]{Gelfand09}
{Gelfand}, J.~D., {Slane}, P.~O., \& {Zhang}, W. 2009, ApJ, 703, 2051

\bibitem[{{Geppert} {et~al.}(1999){Geppert}, {Page}, \&
  {Zannias}}]{1999A&A...345..847G}
{Geppert}, U., {Page}, D., \& {Zannias}, T. 1999, \aap, 345, 847

\bibitem[{{Graham} {et~al.}(2014){Graham}, {Zheng}, {Filippenko}, {Challis},
  {Kirshner}, {Parrent}, {Castander}, {Casas}, {Garcia-Alvarez},
  {Perez-Valladares}, {Miquel}, {Smith}, {Schubnell}, {Kessler}, {Scolnic},
  {Covarrubias}, {Wolf}, {Fischer}, {Fischer}, {Gladney}, {March}, {Sako},
  {Brown}, {Krisciunas}, {Suntzeff}, {D'Andrea}, {Nichol}, {Papadopoulos},
  {Smith}, {Sullivan}, {Maartens}, {Gupta}, {Kovacs}, {Kuhlmann}, {Spinka},
  {Ahn}, {Finley}, {Frieman}, {Marriner}, {Wester}, {Aldering}, {Kim},
  {Thomas}, {Barbary}, {Bloom}, {Goldstein}, {Nugent}, {Perlmutter}, {Foley},
  {Desai}, \& {Paech}}]{2014ATel.6635....1G}
{Graham}, M.~L., {Zheng}, W., {Filippenko}, A.~V., {et~al.} 2014, The
  Astronomer's Telegram, 6635

\bibitem[{{Guillochon} {et~al.}(2017){Guillochon}, {Parrent}, {Kelley}, \&
  {Margutti}}]{2017ApJ...835...64G}
{Guillochon}, J., {Parrent}, J., {Kelley}, L.~Z., \& {Margutti}, R. 2017, \apj,
  835, 64

\bibitem[{{G{\"u}neyda{\c s}} \& {Ek{\c s}i}(2013)}]{2013MNRAS.430L..59G}
{G{\"u}neyda{\c s}}, A. \& {Ek{\c s}i}, K.~Y. 2013, \mnras, 430, L59

\bibitem[{{Hinshaw} {et~al.}(2013){Hinshaw}, {Larson}, {Komatsu}, {Spergel},
  {Bennett}, {Dunkley}, {Nolta}, {Halpern}, {Hill}, {Odegard}, {Page}, {Smith},
  {Weiland}, {Gold}, {Jarosik}, {Kogut}, {Limon}, {Meyer}, {Tucker}, {Wollack},
  \& {Wright}}]{2013ApJS..208...19H}
{Hinshaw}, G., {Larson}, D., {Komatsu}, E., {et~al.} 2013, \apjs, 208, 19

\bibitem[{{Ho}(2011)}]{2011MNRAS.414.2567H}
{Ho}, W.~C.~G. 2011, \mnras, 414, 2567

\bibitem[{{Inserra} {et~al.}(2013{\natexlab{a}}){Inserra}, {Smartt}, {Fraser},
  {Young}, {Smith}, {Wright}, {Kotak}, {McCrum}, {Magill}, {Chen},
  {Pastorello}, {Benetti}, {Valenti}, {Bresolin}, {Kudritzki}, {Tonry},
  {Magnier}, {Huber}, {Chambers}, {Kaiser}, {Morgan}, {Burgett}, {Heasley},
  {Sweeney}, {Waters}, {Flewelling}, {Stubbs}, {Price}, {Sollerman}, {Taddia},
  {Ergon}, {Leloudas}, {Taubenberger}, \& {Mahabal}}]{2013CBET.3467....3I}
{Inserra}, C., {Smartt}, S.~J., {Fraser}, M., {et~al.} 2013{\natexlab{a}},
  Central Bureau Electronic Telegrams, 3467

\bibitem[{{Inserra} {et~al.}(2013{\natexlab{b}}){Inserra}, {Smartt}, {Fraser},
  {Young}, {Smith}, {Wright}, {Kotak}, {McCrum}, {Magill}, {Chen},
  {Pastorello}, {Benetti}, {Valenti}, {Bresolin}, {Kudritzki}, {Tonry},
  {Magnier}, {Huber}, {Chambers}, {Kaiser}, {Morgan}, {Burgett}, {Heasley},
  {Sweeney}, {Waters}, {Flewelling}, {Stubbs}, {Price}, {Sollerman}, {Taddia},
  {Ergon}, {Leloudas}, \& {Taubenberger}}]{2013CBET.3463....2I}
{Inserra}, C., {Smartt}, S.~J., {Fraser}, M., {et~al.} 2013{\natexlab{b}},
  Central Bureau Electronic Telegrams, 3463

\bibitem[{{Kaiser} {et~al.}(2010){Kaiser}, {Burgett}, {Chambers}, {Denneau},
  {Heasley}, {Jedicke}, {Magnier}, {Morgan}, {Onaka}, \&
  {Tonry}}]{2010SPIE.7733E..0EK}
{Kaiser}, N., {Burgett}, W., {Chambers}, K., {et~al.} 2010, in \procspie, Vol.
  7733, Ground-based and Airborne Telescopes III, 77330E

\bibitem[{{Kasen} \& {Bildsten}(2010)}]{Kasen10}
{Kasen}, D. \& {Bildsten}, L. 2010, \apj, 717, 245

\bibitem[{{Katz} {et~al.}(2012){Katz}, {Sapir}, \& {Waxman}}]{Katz12}
{Katz}, B., {Sapir}, N., \& {Waxman}, E. 2012, ApJ, 747, 147

\bibitem[{{Kirk} {et~al.}(2009){Kirk}, {Lyubarsky}, \& {Petri}}]{Kirk09}
{Kirk}, J.~G., {Lyubarsky}, Y., \& {Petri}, J. 2009, in Astrophysics and Space
  Science Library, Vol. 357, Astrophysics and Space Science Library, ed.
  W.~{Becker}, 421

\bibitem[{{Kotera} {et~al.}(2013){Kotera}, {Phinney}, \& {Olinto}}]{KPO13}
{Kotera}, K., {Phinney}, E.~S., \& {Olinto}, A.~V. 2013, MNRAS, 432, 3228

\bibitem[{{Le Guillou} {et~al.}(2015){Le Guillou}, {Mitra}, {Baumont},
  {Chotard}, {Leget}, {Fraser}, {Galbany}, {Dennefeld}, {Inserra}, {Maguire},
  {Smartt}, {Smith}, {Sullivan}, {Valenti}, {Yaron}, {Young}, {Manulis},
  {Baltay}, {Ellman}, {Hadjiyska}, {McKinnon}, {Rabinowitz}, {Rostami},
  {Feindt}, {Kowalski}, {Nugent}, {Wright}, {Chambers}, {Flewelling}, {Huber},
  {Magnier}, {Tonry}, {Waters}, \& {Wainscoat}}]{2015ATel.7102....1L}
{Le Guillou}, L., {Mitra}, A., {Baumont}, S., {et~al.} 2015, The Astronomer's
  Telegram, 7102

\bibitem[{{Leget} {et~al.}(2014){Leget}, {Guillou}, {Fleury}, {Baumont},
  {Balland}, {Botticella}, {Elias-Rosa}, {Pastorello}, {Benetti}, {Inserra},
  {Smartt}, {Smith}, {Young}, {Sullivan}, {Taubenberger}, {Valenti}, {Fraser},
  {Yaron}, {Manulis}, {Gal-Yam}, {Knapic}, {Smareglia}, {Molinaro}, {Baltay},
  {Ellman}, {Hadjiyska}, {McKinnon}, {Rabinowitz}, {Walker}, {Feindt},
  {Kowalski}, {Nugent}, \& {Wyrzykowski}}]{2014ATel.5718....1L}
{Leget}, P.-F., {Guillou}, L.~L., {Fleury}, M., {et~al.} 2014, The Astronomer's
  Telegram, 5718

\bibitem[{{Leloudas} {et~al.}(2014){Leloudas}, {Ergon}, {Taddia}, {Nyholm},
  {Sollerman}, {Inserra}, {Scalzo}, {Benetti}, {Pastorello}, {Smartt}, {Smith},
  {Young}, {Sullivan}, {Taubenberger}, {Valenti}, {Fraser}, {Yaron}, {Gal-Yam},
  {Manulis}, {Knapic}, {Smareglia}, {Molinaro}, {Baltay}, {Ellman},
  {Hadjiyska}, {McKinnon}, {Rabinowitz}, {Walker}, {Feindt}, {Kowalski}, \&
  {Nugent}}]{2014ATel.5839....1L}
{Leloudas}, G., {Ergon}, M., {Taddia}, F., {et~al.} 2014, The Astronomer's
  Telegram, 5839

\bibitem[{{Liu} {et~al.}(2017){Liu}, {Modjaz}, \&
  {Bianco}}]{2016arXiv161207321L}
{Liu}, Y.-Q., {Modjaz}, M., \& {Bianco}, F.~B. 2017, \apj, 845, 85

\bibitem[{{Lorenz} {et~al.}(1993){Lorenz}, {Ravenhall}, \&
  {Pethick}}]{1993PhRvL..70..379L}
{Lorenz}, C.~P., {Ravenhall}, D.~G., \& {Pethick}, C.~J. 1993, Physical Review
  Letters, 70, 379

\bibitem[{{Lunnan} {et~al.}(2014){Lunnan}, {Chornock}, {Berger}, {Laskar},
  {Fong}, {Rest}, {Sanders}, {Challis}, {Drout}, {Foley}, {Huber}, {Kirshner},
  {Leibler}, {Marion}, {McCrum}, {Milisavljevic}, {Narayan}, {Scolnic},
  {Smartt}, {Smith}, {Soderberg}, {Tonry}, {Burgett}, {Chambers}, {Flewelling},
  {Hodapp}, {Kaiser}, {Magnier}, {Price}, \& {Wainscoat}}]{2014ApJ...787..138L}
{Lunnan}, R., {Chornock}, R., {Berger}, E., {et~al.} 2014, \apj, 787, 138

\bibitem[{{Lunnan} {et~al.}(2013){Lunnan}, {Chornock}, {Berger},
  {Milisavljevic}, {Drout}, {Sanders}, {Challis}, {Czekala}, {Foley}, {Fong},
  {Huber}, {Kirshner}, {Leibler}, {Marion}, {McCrum}, {Narayan}, {Rest},
  {Roth}, {Scolnic}, {Smartt}, {Smith}, {Soderberg}, {Stubbs}, {Tonry},
  {Burgett}, {Chambers}, {Kudritzki}, {Magnier}, \&
  {Price}}]{2013ApJ...771...97L}
{Lunnan}, R., {Chornock}, R., {Berger}, E., {et~al.} 2013, \apj, 771, 97

\bibitem[{{Lunnan} {et~al.}(2016){Lunnan}, {Chornock}, {Berger},
  {Milisavljevic}, {Jones}, {Rest}, {Fong}, {Fransson}, {Margutti}, {Drout},
  {Blanchard}, {Challis}, {Cowperthwaite}, {Foley}, {Kirshner}, {Morrell},
  {Riess}, {Roth}, {Scolnic}, {Smartt}, {Smith}, {Villar}, {Chambers},
  {Draper}, {Huber}, {Kaiser}, {Kudritzki}, {Magnier}, {Metcalfe}, \&
  {Waters}}]{2016ApJ...831..144L}
{Lunnan}, R., {Chornock}, R., {Berger}, E., {et~al.} 2016, \apj, 831, 144

\bibitem[{{McCrum} {et~al.}(2014){McCrum}, {Smartt}, {Kotak}, {Rest},
  {Jerkstrand}, {Inserra}, {Rodney}, {Chen}, {Howell}, {Huber}, {Pastorello},
  {Tonry}, {Bresolin}, {Kudritzki}, {Chornock}, {Berger}, {Smith},
  {Botticella}, {Foley}, {Fraser}, {Milisavljevic}, {Nicholl}, {Riess},
  {Stubbs}, {Valenti}, {Wood-Vasey}, {Wright}, {Young}, {Drout}, {Czekala},
  {Burgett}, {Chambers}, {Draper}, {Flewelling}, {Hodapp}, {Kaiser}, {Magnier},
  {Metcalfe}, {Price}, {Sweeney}, \& {Wainscoat}}]{2014MNRAS.437..656M}
{McCrum}, M., {Smartt}, S.~J., {Kotak}, R., {et~al.} 2014, \mnras, 437, 656

\bibitem[{{McCrum} {et~al.}(2015){McCrum}, {Smartt}, {Rest}, {Smith}, {Kotak},
  {Rodney}, {Young}, {Chornock}, {Berger}, {Foley}, {Fraser}, {Wright},
  {Scolnic}, {Tonry}, {Urata}, {Huang}, {Pastorello}, {Botticella}, {Valenti},
  {Mattila}, {Kankare}, {Farrow}, {Huber}, {Stubbs}, {Kirshner}, {Bresolin},
  {Burgett}, {Chambers}, {Draper}, {Flewelling}, {Jedicke}, {Kaiser},
  {Magnier}, {Metcalfe}, {Morgan}, {Price}, {Sweeney}, {Wainscoat}, \&
  {Waters}}]{2015MNRAS.448.1206M}
{McCrum}, M., {Smartt}, S.~J., {Rest}, A., {et~al.} 2015, \mnras, 448, 1206

\bibitem[{{Metzger} {et~al.}(2014){Metzger}, {Vurm}, {Hasco{\"e}t}, \&
  {Beloborodov}}]{Metzger14}
{Metzger}, B.~D., {Vurm}, I., {Hasco{\"e}t}, R., \& {Beloborodov}, A.~M. 2014,
  MNRAS, 437, 703

\bibitem[{{Miller} {et~al.}(2009){Miller}, {Chornock}, {Perley},
  {Ganeshalingam}, {Li}, {Butler}, {Bloom}, {Smith}, {Modjaz}, {Poznanski},
  {Filippenko}, {Griffith}, {Shiode}, \& {Silverman}}]{Miller09}
{Miller}, A.~A., {Chornock}, R., {Perley}, D.~A., {et~al.} 2009, ApJ, 690, 1303

\bibitem[{{Murase} {et~al.}(2015){Murase}, {Kashiyama}, {Kiuchi}, \&
  {Bartos}}]{Murase15}
{Murase}, K., {Kashiyama}, K., {Kiuchi}, K., \& {Bartos}, I. 2015, ApJ, 805, 82

\bibitem[{{Murase} {et~al.}(2011){Murase}, {Thompson}, {Lacki}, \&
  {Beacom}}]{Murase11}
{Murase}, K., {Thompson}, T.~A., {Lacki}, B.~C., \& {Beacom}, J.~F. 2011, Phys.
  Rev. D, 84, 043003

\bibitem[{{Murase} {et~al.}(2014){Murase}, {Thompson}, \& {Ofek}}]{Murase14}
{Murase}, K., {Thompson}, T.~A., \& {Ofek}, E.~O. 2014, MNRAS, 440, 2528

\bibitem[{{Muslimov} \& {Page}(1995)}]{1995ApJ...440L..77M}
{Muslimov}, A. \& {Page}, D. 1995, \apjl, 440, L77

\bibitem[{{Nicholl} {et~al.}(2014){Nicholl}, {Smartt}, {Jerkstrand}, {Inserra},
  {Anderson}, {Baltay}, {Benetti}, {Chen}, {Elias-Rosa}, {Feindt}, {Fraser},
  {Gal-Yam}, {Hadjiyska}, {Howell}, {Kotak}, {Lawrence}, {Leloudas},
  {Margheim}, {Mattila}, {McCrum}, {McKinnon}, {Mead}, {Nugent}, {Rabinowitz},
  {Rest}, {Smith}, {Sollerman}, {Sullivan}, {Taddia}, {Valenti}, {Walker}, \&
  {Young}}]{Nicholl14}
{Nicholl}, M., {Smartt}, S.~J., {Jerkstrand}, A., {et~al.} 2014, MNRAS, 444,
  2096

\bibitem[{{Ofek} {et~al.}(2007){Ofek}, {Cameron}, {Kasliwal}, {Gal-Yam}, {Rau},
  {Kulkarni}, {Frail}, {Chandra}, {Cenko}, {Soderberg}, \& {Immler}}]{Ofek07}
{Ofek}, E.~O., {Cameron}, P.~B., {Kasliwal}, M.~M., {et~al.} 2007, ApJ Lett.,
  659, L13

\bibitem[{{Papadopoulos} {et~al.}(2013){Papadopoulos}, {Sullivan}, {D'Andrea},
  {Nichol}, {Maguire}, {Kessler}, {Covarrubias}, {Cane}, {Fischer}, {Gladney},
  {March}, {Sako}, {Brown}, {Krisciunas}, {Suntzeff}, {Maartens}, {Smith},
  {Barbary}, {Bernstein}, {Biswas}, {Gupta}, {Kovacs}, {Kuhlmann}, {Spinka},
  {Ahn}, {Finley}, {Frieman}, {Marriner}, {Wester}, {Aldering}, {Bloom},
  {Goldstein}, {Kim}, {Nugent}, {Perlmutter}, {Thomas}, {Foley}, {Desai},
  {Paech}, {Smith}, \& {Schubnell}}]{2013ATel.5603....1P}
{Papadopoulos}, A., {Sullivan}, M., {D'Andrea}, C., {et~al.} 2013, The
  Astronomer's Telegram, 5603

\bibitem[{{Pastorello} {et~al.}(2010){Pastorello}, {Smartt}, {Botticella},
  {Magill}, {Young}, {Valenti}, \& {Kotak}}]{2010CBET.2413....1P}
{Pastorello}, A., {Smartt}, S.~J., {Botticella}, M.~T., {et~al.} 2010, Central
  Bureau Electronic Telegrams, 2413

\bibitem[{{Pignata} {et~al.}(2013){Pignata}, {Apostolovski}, {Paillas},
  {Varela}, {Bufano}, {Olivares}, {Takats}, {Hamuy}, {Antezana}, {Gonzalez},
  {Cartier}, {Forster}, {Silva}, {Carrasco}, {Ramirez}, {Aros}, {Conuel},
  {Folatelli}, {Reichart}, {Haislip}, {Moore}, {LaCluyze}, {Inserra},
  {Kankare}, {Kangas}, {Mattila}, {Fraser}, {Scalzo}, {Nicholl}, {Gal-Yam},
  {Yaron}, {Benetti}, {Pastorello}, {Valenti}, {Taubenberger}, {Smartt},
  {Smith}, {Young}, {Sullivan}, {Knapic}, {Molinaro}, \&
  {Smareglia}}]{2013CBET.3644....1P}
{Pignata}, G., {Apostolovski}, Y., {Paillas}, E., {et~al.} 2013, Central Bureau
  Electronic Telegrams, 3644

\bibitem[{{Prajs} {et~al.}(2015){Prajs}, {Cartier}, {Frohmaier}, {De Cia},
  {Galbany}, {Inserra}, {Maguire}, {Smartt}, {Smith}, {Sullivan}, {Valenti},
  {Yaron}, {Young}, {Manulis}, {Baltay}, {Ellman}, {Hadjiyska}, {McKinnon},
  {Rabinowitz}, {Rostami}, {Feindt}, {Kowalski}, \&
  {Nugent}}]{2015ATel.7412....1P}
{Prajs}, S., {Cartier}, R., {Frohmaier}, C., {et~al.} 2015, The Astronomer's
  Telegram, 7412

\bibitem[{{Quimby} {et~al.}(2010{\natexlab{a}}){Quimby}, {Gal-Yam}, {Arcavi},
  {Ben-Ami}, {Xu}, {Sternberg}, {Kasliwal}, {Ofek}, {Kulkarni}, {Nugent},
  {Filippenko}, \& {Cenko}}]{2010ATel.2634....1Q}
{Quimby}, R., {Gal-Yam}, A., {Arcavi}, I., {et~al.} 2010{\natexlab{a}}, The
  Astronomer's Telegram, 2634

\bibitem[{{Quimby}(2012)}]{Quimby12}
{Quimby}, R.~M. 2012, in IAU Symposium, Vol. 279, IAU Symposium, 22--28

\bibitem[{{Quimby} {et~al.}(2012){Quimby}, {Arcavi}, {Sternberg}, {Ben-Ami},
  {Yaron}, {Gal-Yam}, {Graham}, {Cenko}, {Filippenko}, {Perley}, {Cao}, \&
  {Kulkarni}}]{2012ATel.4121....1Q}
{Quimby}, R.~M., {Arcavi}, I., {Sternberg}, A., {et~al.} 2012, The Astronomer's
  Telegram, 4121

\bibitem[{{Quimby} {et~al.}(2011{\natexlab{a}}){Quimby}, {Cenko}, {Yaron},
  {Xu}, {Horesh}, {Silverman}, {Filippenko}, {Derose}, \&
  {Nugent}}]{2011ATel.3465....1Q}
{Quimby}, R.~M., {Cenko}, S.~B., {Yaron}, O., {et~al.} 2011{\natexlab{a}}, The
  Astronomer's Telegram, 3465

\bibitem[{{Quimby} {et~al.}(2011{\natexlab{b}}){Quimby}, {Gal-Yam}, {Arcavi},
  {Yaron}, {Horesh}, \& {Mooley}}]{2011ATel.3841....1Q}
{Quimby}, R.~M., {Gal-Yam}, A., {Arcavi}, I., {et~al.} 2011{\natexlab{b}}, The
  Astronomer's Telegram, 3841

\bibitem[{{Quimby} {et~al.}(2013{\natexlab{a}}){Quimby}, {Kulkarni}, {Ofek},
  {Kasliwal}, {Gal-Yam}, {Arcavi}, {Ben-Ami}, {Xu}, {Sternberg}, {Silverman},
  {Cenko}, {Kleiser}, {Nugent}, {Howell}, {Inserra}, {Smartt}, {Fraser},
  {Young}, {Smith}, {Wright}, {Kotak}, {McCrum}, {Magill}, {Chen},
  {Pastorello}, {Benetti}, {Valenti}, {Bresolin}, {Kudritzki}, {Tonry},
  {Magnier}, {Huber}, {Chambers}, {Kaiser}, {Morgan}, {Burgett}, {Heasley},
  {Sweeney}, {Waters}, {Flewelling}, {Stubbs}, {Price}, {Sollerman}, {Taddia},
  {Ergon}, {Leloudas}, \& {Taubenberger}}]{2013CBET.3461....1Q}
{Quimby}, R.~M., {Kulkarni}, S., {Ofek}, E., {et~al.} 2013{\natexlab{a}},
  Central Bureau Electronic Telegrams, 3461

\bibitem[{{Quimby} {et~al.}(2010{\natexlab{b}}){Quimby}, {Kulkarni}, {Ofek},
  {Kasliwal}, {Gal-Yam}, {Ben-Ami}, {Badenes}, {Sternberg}, {Botyanszki},
  {Nugent}, \& {Howell}}]{2010ATel.2979....1Q}
{Quimby}, R.~M., {Kulkarni}, S., {Ofek}, E., {et~al.} 2010{\natexlab{b}}, The
  Astronomer's Telegram, 2979

\bibitem[{{Quimby} {et~al.}(2011{\natexlab{c}}){Quimby}, {Kulkarni},
  {Kasliwal}, {Gal-Yam}, {Arcavi}, {Sullivan}, {Nugent}, {Thomas}, {Howell},
  {Nakar}, {Bildsten}, {Theissen}, {Law}, {Dekany}, {Rahmer}, {Hale}, {Smith},
  {Ofek}, {Zolkower}, {Velur}, {Walters}, {Henning}, {Bui}, {McKenna},
  {Poznanski}, {Cenko}, \& {Levitan}}]{Quimby11}
{Quimby}, R.~M., {Kulkarni}, S.~R., {Kasliwal}, M.~M., {et~al.}
  2011{\natexlab{c}}, Nature, 474, 487

\bibitem[{{Quimby} {et~al.}(2011{\natexlab{d}}){Quimby}, {Kulkarni},
  {Kasliwal}, {Gal-Yam}, {Arcavi}, {Sullivan}, {Nugent}, {Thomas}, {Howell},
  {Nakar}, {Bildsten}, {Theissen}, {Law}, {Dekany}, {Rahmer}, {Hale}, {Smith},
  {Ofek}, {Zolkower}, {Velur}, {Walters}, {Henning}, {Bui}, {McKenna},
  {Poznanski}, {Cenko}, \& {Levitan}}]{2011Natur.474..487Q}
{Quimby}, R.~M., {Kulkarni}, S.~R., {Kasliwal}, M.~M., {et~al.}
  2011{\natexlab{d}}, \nat, 474, 487

\bibitem[{{Quimby} {et~al.}(2013{\natexlab{b}}){Quimby}, {Yuan}, {Akerlof}, \&
  {Wheeler}}]{Quimby13}
{Quimby}, R.~M., {Yuan}, F., {Akerlof}, C., \& {Wheeler}, J.~C.
  2013{\natexlab{b}}, MNRAS, 431, 912

\bibitem[{{Rau} {et~al.}(2009){Rau}, {Kulkarni}, {Law}, {Bloom}, {Ciardi},
  {Djorgovski}, {Fox}, {Gal-Yam}, {Grillmair}, {Kasliwal}, {Nugent}, {Ofek},
  {Quimby}, {Reach}, {Shara}, {Bildsten}, {Cenko}, {Drake}, {Filippenko},
  {Helfand}, {Helou}, {Howell}, {Poznanski}, \&
  {Sullivan}}]{2009PASP..121.1334R}
{Rau}, A., {Kulkarni}, S.~R., {Law}, N.~M., {et~al.} 2009, \pasp, 121, 1334

\bibitem[{{Renault-Tinacci} {et~al.}(2015){Renault-Tinacci}, {Grenier}, \&
  {Harding}}]{2015ICRC...34..843R}
{Renault-Tinacci}, N., {Grenier}, I., \& {Harding}, A.~K. 2015, in
  International Cosmic Ray Conference, Vol.~34, 34th International Cosmic Ray
  Conference (ICRC2015), 843

\bibitem[{{Scalzo} {et~al.}(2014){Scalzo}, {Yuan}, {Childress}, {Schmidt},
  {Tucker}, {Inserra}, {Smartt}, {Nicholl}, {Fraser}, {Benetti}, {Pastorello},
  {Smith}, {Young}, {Sullivan}, {Taubenberger}, {Valenti}, {Yaron}, {Gal-Yam},
  {Knapic}, {Smareglia}, \& {Molinaro}}]{2014CBET.3836....1S}
{Scalzo}, R., {Yuan}, F., {Childress}, M., {et~al.} 2014, Central Bureau
  Electronic Telegrams, 3836

\bibitem[{{Shapiro} \& {Teukolsky}(1983)}]{Shapiro83}
{Shapiro}, S.~L. \& {Teukolsky}, S.~A. 1983, {Black holes, white dwarfs, and
  neutron stars: The physics of compact objects} (John Wiley and Son. Inc.,
  Hoboken, NJ)

\bibitem[{{Smartt} {et~al.}(2012){Smartt}, {Inserra}, {Fraser}, {Wright},
  {Young}, {Pastorello}, {Valenti}, {Taubenberger}, {Sullivan}, {Gal-Yam},
  {Yaron}, {Pignata}, {Howerton}, {Baltay}, {Ellman}, {Hadjiyska}, {McKinnon},
  {Rabinowitz}, {Feindt}, {Kowalski}, \& {Nugent}}]{2012ATel.4299....1S}
{Smartt}, S.~J., {Inserra}, C., {Fraser}, M., {et~al.} 2012, The Astronomer's
  Telegram, 4299

\bibitem[{{Smith} {et~al.}(2016){Smith}, {Sullivan}, {D'Andrea}, {Castander},
  {Casas}, {Prajs}, {Papadopoulos}, {Nichol}, {Karpenka}, {Bernard}, {Brown},
  {Cartier}, {Cooke}, {Curtin}, {Davis}, {Finley}, {Foley}, {Gal-Yam},
  {Goldstein}, {Gonz{\'a}lez-Gait{\'a}n}, {Gupta}, {Howell}, {Inserra},
  {Kessler}, {Lidman}, {Marriner}, {Nugent}, {Pritchard}, {Sako}, {Smartt},
  {Smith}, {Spinka}, {Thomas}, {Wolf}, {Zenteno}, {Abbott}, {Benoit-L{\'e}vy},
  {Bertin}, {Brooks}, {Buckley-Geer}, {Carnero Rosell}, {Carrasco Kind},
  {Carretero}, {Crocce}, {Cunha}, {da Costa}, {Desai}, {Diehl}, {Doel},
  {Estrada}, {Evrard}, {Flaugher}, {Fosalba}, {Frieman}, {Gerdes}, {Gruen},
  {Gruendl}, {James}, {Kuehn}, {Kuropatkin}, {Lahav}, {Li}, {Marshall},
  {Martini}, {Miller}, {Miquel}, {Nord}, {Ogando}, {Plazas}, {Reil}, {Romer},
  {Roodman}, {Rykoff}, {Sanchez}, {Scarpine}, {Schubnell}, {Sevilla-Noarbe},
  {Soares-Santos}, {Sobreira}, {Suchyta}, {Swanson}, {Tarle}, {Walker},
  {Wester}, \& {DES Collaboration}}]{2016ApJ...818L...8S}
{Smith}, M., {Sullivan}, M., {D'Andrea}, C.~B., {et~al.} 2016, \apjl, 818, L8

\bibitem[{{Smith} {et~al.}(2008){Smith}, {Chornock}, {Li}, {Ganeshalingam},
  {Silverman}, {Foley}, {Filippenko}, \& {Barth}}]{Smith08}
{Smith}, N., {Chornock}, R., {Li}, W., {et~al.} 2008, ApJ, 686, 467

\bibitem[{{Smith} {et~al.}(2010){Smith}, {Chornock}, {Silverman}, {Filippenko},
  \& {Foley}}]{Smith10}
{Smith}, N., {Chornock}, R., {Silverman}, J.~M., {Filippenko}, A.~V., \&
  {Foley}, R.~J. 2010, \apj, 709, 856

\bibitem[{{Smith} \& {McCray}(2007)}]{Smith07}
{Smith}, N. \& {McCray}, R. 2007, ApJ Lett., 671, L17

\bibitem[{{Suzuki} \& {Maeda}(2017)}]{Suzuki16}
{Suzuki}, A. \& {Maeda}, K. 2017, \mnras, 466, 2633

\bibitem[{{Tanaka} \& {Takahara}(2011)}]{Tanaka11}
{Tanaka}, S.~J. \& {Takahara}, F. 2011, ApJ, 741, 40

\bibitem[{{Tomasella} {et~al.}(2012){Tomasella}, {Benetti}, {Pastorello},
  {Cappellaro}, {Turatto}, \& {Harutyunyan}}]{2012ATel.4512....1T}
{Tomasella}, L., {Benetti}, S., {Pastorello}, A., {et~al.} 2012, The
  Astronomer's Telegram, 4512

\bibitem[{{Torres-Forn{\'e}} {et~al.}(2016){Torres-Forn{\'e}},
  {Cerd{\'a}-Dur{\'a}n}, {Pons}, \& {Font}}]{2016MNRAS.456.3813T}
{Torres-Forn{\'e}}, A., {Cerd{\'a}-Dur{\'a}n}, P., {Pons}, J.~A., \& {Font},
  J.~A. 2016, \mnras, 456, 3813

\bibitem[{{Vigan{\`o}} \& {Pons}(2012)}]{2012MNRAS.425.2487V}
{Vigan{\`o}}, D. \& {Pons}, J.~A. 2012, \mnras, 425, 2487

\bibitem[{{Vinko} {et~al.}(2010){Vinko}, {Zheng}, {Romadan}, {Quimby},
  {Whallon}, {Pandey}, {Fang}, {Akerlof}, {Pasque}, {Verkinderen}, {Wheeler},
  {Chatzopoulos}, \& {Caldwell}}]{2010CBET.2556....1V}
{Vinko}, J., {Zheng}, W., {Romadan}, A., {et~al.} 2010, Central Bureau
  Electronic Telegrams, 2556

\bibitem[{{Vreeswijk} {et~al.}(2014){Vreeswijk}, {Savaglio}, {Gal-Yam}, {De
  Cia}, {Quimby}, {Sullivan}, {Cenko}, {Perley}, {Filippenko}, {Clubb},
  {Taddia}, {Sollerman}, {Leloudas}, {Arcavi}, {Rubin}, {Kasliwal}, {Cao},
  {Yaron}, {Tal}, {Ofek}, {Capone}, {Kutyrev}, {Toy}, {Nugent}, {Laher},
  {Surace}, \& {Kulkarni}}]{2014ApJ...797...24V}
{Vreeswijk}, P.~M., {Savaglio}, S., {Gal-Yam}, A., {et~al.} 2014, \apj, 797, 24

\bibitem[{{Wright} {et~al.}(2012){Wright}, {Cellier-Holzem}, {Inserra},
  {Smartt}, {Fraser}, {Young}, {Smith}, {Pastorello}, {Valenti}, {Benetti},
  {Taubenberger}, {Sullivan}, {Gal-Yam}, {Yaron}, \&
  {Arcavi}}]{2012ATel.4313....1W}
{Wright}, D., {Cellier-Holzem}, F., {Inserra}, C., {et~al.} 2012, The
  Astronomer's Telegram, 4313

\bibitem[{{Yan} {et~al.}(2015){Yan}, {Quimby}, {Ofek}, {Gal-Yam}, {Mazzali},
  {Perley}, {Vreeswijk}, {Leloudas}, {de Cia}, {Masci}, {Cenko}, {Cao},
  {Kulkarni}, {Nugent}, {Rebbapragada}, {Wo{\'z}niak}, \&
  {Yaron}}]{2015ApJ...814..108Y}
{Yan}, L., {Quimby}, R., {Ofek}, E., {et~al.} 2015, \apj, 814, 108

\end{thebibliography}
\clearpage

\begin{appendix}
\onecolumn
\section{}

\begin{table}[!h] 
\centering 
\tiny
\caption{Equatorial and Galactic coordinates, redshift, and SN date in Gregorian and MJD calendars of the studied SLSNe. Date refers either to the peak date indicated by xx$^{\rm{p}}$ or to the discovery date indicated by xx$^{\rm{d}}$.} 
\label{tab_studied_SNe} 
%\resizebox{\textwidth}{!}{
\begin{tabular}{lcccccccl} 
Name & ra [deg] & dec [deg] & $l$ [deg] & $b$ [deg] & Redshift & Date & Date [MJD] & References \\ 
\hline 
\noalign{\smallskip} 
\object{SN2008fz} & 349.07 & 11.71 & 89.10 & -44.82 & 0.1330 & 2008-Sep-22$^{\rm{d}}$ & 54731 & \cite{2010ApJ...718L.127D} \\ 
\noalign{\smallskip} 
\object{SN2009jh} & 222.29 & 29.42 & 44.97 & 64.05 & 0.3490 & 2009-Aug-2$^{\rm{d}}$ & 55045 & \cite{2009CBET.1958....1D} \\ 
\noalign{\smallskip} 
\object{PTF09atu} & 247.60 & 23.64 & 41.73 & 40.80 & 0.5010 & 2009-Aug-17$^{\rm{p}}$ & 55060 & \cite{2011Natur.474..487Q} \\ 
\noalign{\smallskip} 
\object{PTF09cnd} & 243.04 & 51.49 & 80.03 & 45.37 & 0.2580 & 2009-Sep-6$^{\rm{p}}$ & 55080 & \cite{2009ATel.2241....1C} \\ 
\noalign{\smallskip} 
\object{CSS100217} & 157.30 & 40.71 & 178.77 & 57.81 & 0.1470 & 2010-Feb-17$^{\rm{d}}$ & 55244 & \cite{2010ATel.2544....1D} \\ 
\noalign{\smallskip} 
\object{PS1-10pm} & 183.18 & 46.99 & 141.37 & 68.73 & 1.2060 & 2010-Feb-24$^{\rm{d}}$ & 55248 & \cite{2015MNRAS.448.1206M} \\ 
\noalign{\smallskip} 
\object{SN2010gx} & 171.44 & -8.83 & 269.96 & 48.48 & 0.2300 & 2010-Mar-13$^{\rm{d}}$ & 55268 & \cite{2010CBET.2413....1P} \\ 
\noalign{\smallskip} 
\object{PTF10heh} & 192.22 & 13.44 & 300.29 & 76.30 & 0.3380 & 2010-Apr-4$^{\rm{d}}$ & 55290.3 & \cite{2010ATel.2634....1Q} \\ 
\noalign{\smallskip} 
\object{PTF10hgi} & 249.45 & 6.21 & 22.46 & 32.45 & 0.1000 & 2010-May-10$^{\rm{p}}$ & 55326.4 & \cite{2013CBET.3461....1Q} \\ 
\noalign{\smallskip} 
\object{PTF10qa} & 353.93 & 10.78 & 94.56 & -47.88 & 0.2840 & 2010-Jun-18$^{\rm{d}}$ & 55365 & \cite{2012Sci...337..927G} \\ 
\noalign{\smallskip} 
\object{PS1-10ky} & 333.41 & 1.24 & 63.24 & -42.62 & 0.9558 & 2010-Jul-20$^{\rm{p}}$ & 55397 & \cite{2011ApJ...743..114C} \\ 
\noalign{\smallskip} 
\object{PS1-10ahf} & 353.12 & -0.36 & 84.36 & -57.20 & 1.1000 & 2010-Aug-6$^{\rm{d}}$ & 55414 & \cite{2015MNRAS.448.1206M} \\ 
\noalign{\smallskip} 
\object{SN2010hy} & 284.89 & 19.41 & 51.19 & 7.00 & 0.1900 & 2010-Sep-4$^{\rm{d}}$ & 55443 & \cite{2010CBET.2461....2C} \\ 
\noalign{\smallskip} 
\object{PTF10vqv} & 45.78 & -1.54 & 179.50 & -49.39 & 0.4520 & 2010-Sep-16$^{\rm{d}}$ & 55455.5 & \cite{2010ATel.2979....1Q} \\ 
\noalign{\smallskip} 
\object{SN2010kd} & 182.00 & 49.23 & 140.80 & 66.37 & 0.1010 & 2010-Nov-14$^{\rm{d}}$ & 55514 & \cite{2010CBET.2556....1V} \\ 
\noalign{\smallskip} 
\object{PS1-10awh} & 333.62 & -0.07 & 62.04 & -43.62 & 0.9084 & 2010-Nov-15$^{\rm{p}}$ & 55515 & \cite{2011ApJ...743..114C} \\ 
\noalign{\smallskip} 
\object{PS1-10bzj} & 52.92 & -27.80 & 223.51 & -54.61 & 0.6490 & 2011-Jan-2$^{\rm{p}}$ & 55563.7 & \cite{2013ApJ...771...97L} \\ 
\noalign{\smallskip} 
\object{PS1-11ap} & 162.12 & 57.15 & 150.32 & 52.94 & 0.5240 & 2011-Feb-21$^{\rm{p}}$ & 55613 & \cite{2014MNRAS.437..656M} \\ 
\noalign{\smallskip} 
\object{PS1-11tt} & 243.19 & 54.07 & 83.53 & 44.64 & 1.2830 & 2011-Apr-24$^{\rm{d}}$ & 55675 & \cite{2014ApJ...787..138L} \\ 
\noalign{\smallskip} 
\object{SN2011ke} & 207.74 & 26.28 & 32.65 & 76.69 & 0.1430 & 2011-May-5$^{\rm{p}}$ & 55686.5 & \cite{2013CBET.3467....3I} \\ 
\noalign{\smallskip} 
\object{PTF11dsf} & 242.89 & 40.30 & 64.01 & 46.97 & 0.3850 & 2011-May-12$^{\rm{d}}$ & 55693 & \cite{2011ATel.3465....1Q} \\ 
\noalign{\smallskip} 
\object{PS1-11afv} & 183.91 & 48.18 & 138.89 & 67.82 & 1.4070 & 2011-May-24$^{\rm{d}}$ & 55705 & \cite{2014ApJ...787..138L} \\ 
\noalign{\smallskip} 
\object{SN2011kf} & 219.24 & 16.52 & 14.91 & 63.38 & 0.2450 & 2011-Dec-30$^{\rm{p}}$ & 55925.5 & \cite{2013CBET.3463....2I} \\ 
\noalign{\smallskip} 
\object{PTF11rks} & 24.94 & 29.92 & 135.25 & -31.79 & 0.1900 & 2012-Jan-6$^{\rm{p}}$ & 55932.7 & \cite{2011ATel.3841....1Q} \\ 
\noalign{\smallskip} 
\object{SN2012il} & 146.55 & 19.84 & 212.45 & 47.15 & 0.1750 & 2012-Jan-15$^{\rm{p}}$ & 55941.4 & \cite{2013CBET.3459....1D} \\ 
\noalign{\smallskip} 
\object{PTF12dam} & 216.19 & 46.23 & 85.12 & 63.46 & 0.1070 & 2012-Jun-10$^{\rm{p}}$ & 56088 & \cite{2012ATel.4121....1Q} \\ 
\noalign{\smallskip} 
\object{LSQ12dlf} & 27.62 & -21.81 & 194.54 & -75.56 & 0.2550 & 2012-Jul-29$^{\rm{p}}$ & 56137.3 & \cite{2012ATel.4299....1S} \\ 
\noalign{\smallskip} 
\object{SSS120810} & 349.51 & -56.16 & 326.50 & -56.49 & 0.1560 & 2012-Aug-8$^{\rm{p}}$ & 56147 & \cite{2012ATel.4313....1W} \\ 
\noalign{\smallskip} 
\object{CSS121015} & 10.68 & 13.47 & 119.69 & -49.34 & 0.2868 & 2012-Oct-25$^{\rm{p}}$ & 56225.5 & \cite{2012ATel.4512....1T} \\ 
\noalign{\smallskip} 
\object{iPTF13ajg} & 249.77 & 37.03 & 59.69 & 41.49 & 0.7403 & 2013-Apr-23$^{\rm{p}}$ & 56405.6 & \cite{2014ApJ...797...24V} \\ 
\noalign{\smallskip} 
\object{SN2013dg} & 199.67 & -7.08 & 314.83 & 55.16 & 0.2650 & 2013-May-17$^{\rm{d}}$ & 56429.7 & \cite{2013CBET.3560....1D} \\ 
\noalign{\smallskip} 
\object{SN2013fc} & 41.29 & -55.74 & 275.20 & -54.83 & 0.0185 & 2013-Aug-20$^{\rm{d}}$ & 56524 & \cite{2013CBET.3644....1P} \\ 
\noalign{\smallskip} 
\object{DES13S2cmm} & 40.64 & -1.36 & 173.57 & -52.94 & 0.6330 & 2013-Sep-24$^{\rm{p}}$ & 56559.2 & \cite{2013ATel.5603....1P} \\ 
\noalign{\smallskip} 
\object{SN2013hx} & 23.89 & -57.96 & 291.83 & -58.20 & 0.1300 & 2013-Dec-31$^{\rm{d}}$ & 56657 & \cite{2014CBET.3836....1S} \\ 
\noalign{\smallskip} 
\object{LSQ14an} & 193.45 & -29.52 & 303.55 & 33.34 & 0.1630 & 2014-Jan-2$^{\rm{d}}$ & 56659 & \cite{2014ATel.5718....1L} \\ 
\noalign{\smallskip} 
\object{iPTF13ehe} & 103.34 & 67.13 & 148.33 & 25.03 & 0.3434 & 2014-Jan-13$^{\rm{p}}$ & 56670.3 & \cite{2015ApJ...814..108Y} \\ 
\noalign{\smallskip} 
\object{LSQ14mo} & 155.67 & -16.92 & 259.30 & 33.07 & 0.2530 & 2014-Jan-30$^{\rm{d}}$ & 56687 & \cite{2014ATel.5839....1L} \\ 
\noalign{\smallskip} 
\object{CSS140222} & 170.15 & 30.47 & 198.00 & 69.86 & 0.0330 & 2014-Feb-22$^{\rm{d}}$ & 56710 & \cite{2014ATel.5934....1B} \\ 
\noalign{\smallskip} 
\object{LSQ14bdq} & 150.42 & -12.37 & 251.17 & 32.95 & 0.3450 & 2014-May-23$^{\rm{p}}$ & 56800 & \cite{2014ATel.6118....1B} \\ 
\noalign{\smallskip} 
\object{PS1-14bj} & 150.54 & 3.66 & 235.53 & 43.29 & 0.5215 & 2014-May-24$^{\rm{p}}$ & 56801.7 & \cite{2016ApJ...831..144L} \\ 
\noalign{\smallskip} 
\object{DES14X2byo} & 35.95 & -6.14 & 173.39 & -59.63 & 0.8690 & 2014-Oct-25$^{\rm{p}}$ & 56955 & \cite{2014ATel.6635....1G} \\ 
\noalign{\smallskip} 
\object{DES14X3taz} & 37.02 & -4.09 & 172.20 & -57.40 & 0.6080 & 2015-Feb-27$^{\rm{p}}$ & 57080 & \cite{2016ApJ...818L...8S} \\ 
\noalign{\smallskip} 
\object{LSQ15abl} & 145.12 & -4.19 & 239.62 & 34.31 & 0.0870 & 2015-Mar-19$^{\rm{d}}$ & 57100 & \cite{2015ATel.7412....1P} \\ 
\noalign{\smallskip} 
\object{SN2015bn} & 173.42 & 0.73 & 264.46 & 57.67 & 0.1136 & 2015-Mar-21$^{\rm{p}}$ & 57102 & \cite{2015ATel.7102....1L} \\ 
\noalign{\smallskip} 
\object{ASASSN-15lh} & 330.56 & -61.66 & 330.07 & -45.55 & 0.2326 & 2015-Jun-5$^{\rm{p}}$ & 57178.5 & \cite{2016Sci...351..257D} \\ 
\hline 
\end{tabular} 
%}
\end{table} 

\begin{table}%[!h]
\centering 
\tiny
\caption{Luminosities from measurements between $t_{\rm peak}$ and $t_{\rm peak} + $3 months, 6 months, 1 year, and 2 years . The second and third (4$^{\rm th}$ and 5$^{\rm th}$, 6$^{\rm th}$ and 7$^{\rm th}$, and 8$^{\rm th}$ and 9$^{\rm th}$, respectively) columns contain the upper limits on the sum of derived luminosities for a 3-month time window (6-month, 1-year, and 2-year, respectively) in the individual energy bands between 600 MeV and 10 GeV, and 600 MeV and 600 GeV, respectively.} 
\label{tab_SNe_fluxes_3m-2y_7bnd} 
\resizebox{\textwidth}{!}{\begin{tabular}{lcccccccc} 
Name & $L_{0.6-10.2\,\rm{GeV}}^{\rm 3\,month}$ & $L_{0.6-600.0\,\rm{GeV}}^{\rm 3\,month}$ & $L_{0.6-10.2\,\rm{GeV}}^{\rm 6\,month}$ & $L_{0.6-600.0\,\rm{GeV}}^{\rm 6\,month}$ & $L_{0.6-10.2\,\rm{GeV}}^{\rm 1\,year}$ & $L_{0.6-600.0\,\rm{GeV}}^{\rm 1\,year}$ & $L_{0.6-10.2\,\rm{GeV}}^{\rm 2\,year}$ & $L_{0.6-600.0\,\rm{GeV}}^{\rm 2\,year}$ \\ 
\noalign{\smallskip} 
 & [erg\,s$^{-1}$] & [erg\,s$^{-1}$] & [erg\,s$^{-1}$] & [erg\,s$^{-1}$] & [erg\,s$^{-1}$] & [erg\,s$^{-1}$] & [erg\,s$^{-1}$] & [erg\,s$^{-1}$]\\ 
\hline 
\noalign{\smallskip} 
\object{SN2008fz} & $<2.8\times10^{44}$ & $<1.3\times10^{45}$ & $<1.5\times10^{44}$ & $<6.2\times10^{45}$ & $<7.6\times10^{43}$ & $<3.0\times10^{45}$ & $<5.0\times10^{43}$ & $<1.5\times10^{45}$ \\ 
\noalign{\smallskip} 
\object{SN2009jh} & $<1.1\times10^{45}$ & $<9.9\times10^{45}$ & $<8.1\times10^{44}$ & $<5.2\times10^{45}$ & $<5.8\times10^{44}$ & $<2.8\times10^{45}$ & $<4.4\times10^{44}$ & $<1.5\times10^{45}$ \\ 
\noalign{\smallskip} 
\object{PTF09atu} & $<3.3\times10^{45}$ & $<2.4\times10^{46}$ & $<2.3\times10^{45}$ & $<1.5\times10^{46}$ & $<1.7\times10^{45}$ & $<7.3\times10^{45}$ & $<1.2\times10^{45}$ & $<3.7\times10^{45}$ \\ 
\noalign{\smallskip} 
\object{PTF09cnd} & $<9.2\times10^{44}$ & $<4.2\times10^{45}$ & $<3.9\times10^{44}$ & $<2.3\times10^{45}$ & $<2.3\times10^{44}$ & $<1.1\times10^{45}$ & $<1.6\times10^{44}$ & $<6.0\times10^{44}$ \\ 
\noalign{\smallskip} 
\object{CSS100217} & $<2.2\times10^{44}$ & $<1.8\times10^{45}$ & $<2.0\times10^{44}$ & $<9.2\times10^{44}$ & $<8.7\times10^{43}$ & $<4.5\times10^{44}$ & $<4.2\times10^{43}$ & $<2.2\times10^{44}$ \\ 
\noalign{\smallskip} 
\object{PS1-10pm} & $<3.0\times10^{46}$ & $<1.8\times10^{47}$ & $<1.3\times10^{46}$ & $<8.7\times10^{46}$ & $<1.3\times10^{46}$ & $<4.7\times10^{46}$ & $<5.0\times10^{45}$ & $<2.3\times10^{46}$ \\ 
\noalign{\smallskip} 
\object{SN2010gx} & $<9.7\times10^{44}$ & $<3.5\times10^{45}$ & $<6.5\times10^{44}$ & $<1.8\times10^{45}$ & $<5.6\times10^{44}$ & $<1.1\times10^{45}$ & $<2.8\times10^{44}$ & $<5.5\times10^{44}$ \\ 
\noalign{\smallskip} 
\object{PTF10heh} & $<2.8\times10^{45}$ & $<1.6\times10^{46}$ & $<1.6\times10^{45}$ & $<7.7\times10^{45}$ & $<5.7\times10^{44}$ & $<3.8\times10^{45}$ & $<5.0\times10^{44}$ & $<2.0\times10^{45}$ \\ 
\noalign{\smallskip} 
\object{PTF10hgi} & $<1.8\times10^{44}$ & $<6.8\times10^{44}$ & $<6.4\times10^{43}$ & $<3.2\times10^{44}$ & $<5.3\times10^{43}$ & $<1.7\times10^{44}$ & $<3.8\times10^{43}$ & $<9.3\times10^{43}$ \\ 
\noalign{\smallskip} 
\object{PTF10qaf} & $<1.1\times10^{45}$ & $<6.8\times10^{45}$ & $<9.1\times10^{44}$ & $<3.5\times10^{45}$ & $<5.4\times10^{44}$ & $<2.5\times10^{45}$ & $<4.0\times10^{44}$ & $<1.3\times10^{45}$ \\ 
\noalign{\smallskip} 
\object{PS1-10ky} & $<1.3\times10^{46}$ & $<1.3\times10^{47}$ & $<9.7\times10^{45}$ & $<6.5\times10^{46}$ & $<7.0\times10^{45}$ & $<3.2\times10^{46}$ & $<5.5\times10^{45}$ & $<1.7\times10^{46}$ \\ 
\noalign{\smallskip} 
\object{PS1-10ahf} & $<2.3\times10^{46}$ & $<2.6\times10^{47}$ & $<1.3\times10^{46}$ & $<1.2\times10^{47}$ & $<1.5\times10^{46}$ & $<6.6\times10^{46}$ & $<7.6\times10^{45}$ & $<3.3\times10^{46}$ \\ 
\noalign{\smallskip} 
\object{SN2010hy} & $<9.6\times10^{44}$ & $<2.5\times10^{45}$ & $<5.4\times10^{44}$ & $<1.3\times10^{45}$ & $<2.8\times10^{44}$ & $<6.4\times10^{44}$ & $<2.5\times10^{44}$ & $<4.1\times10^{44}$ \\ 
\noalign{\smallskip} 
\object{PTF10vqv} & $<3.1\times10^{45}$ & $<2.1\times10^{46}$ & $<2.0\times10^{45}$ & $<1.1\times10^{46}$ & $<1.0\times10^{45}$ & $<5.2\times10^{45}$ & $<9.3\times10^{44}$ & $<2.9\times10^{45}$ \\ 
\noalign{\smallskip} 
\object{SN2010kd} & $<7.7\times10^{43}$ & $<6.8\times10^{44}$ & $<5.3\times10^{43}$ & $<3.5\times10^{44}$ & $<3.5\times10^{43}$ & $<1.7\times10^{44}$ & $<2.0\times10^{43}$ & $<8.7\times10^{43}$ \\ 
\noalign{\smallskip} 
\object{PS1-10awh} & $<1.5\times10^{46}$ & $<1.5\times10^{48}$ & $<9.1\times10^{45}$ & $<8.0\times10^{47}$ & $<7.6\times10^{45}$ & $<4.0\times10^{47}$ & $<4.4\times10^{45}$ & $<1.7\times10^{47}$ \\ 
\noalign{\smallskip} 
\object{PS1-10bzj} & $<8.3\times10^{45}$ & $<4.5\times10^{46}$ & $<4.3\times10^{45}$ & $<2.3\times10^{46}$ & $<3.5\times10^{45}$ & $<1.2\times10^{46}$ & $<2.8\times10^{45}$ & $<6.8\times10^{45}$ \\ 
\noalign{\smallskip} 
\object{PS1-11ap} & $<4.5\times10^{45}$ & $<2.6\times10^{46}$ & $<2.2\times10^{45}$ & $<1.3\times10^{46}$ & $<1.2\times10^{45}$ & $<6.2\times10^{45}$ & $<7.4\times10^{44}$ & $<3.1\times10^{45}$ \\ 
\noalign{\smallskip} 
\object{PS1-11tt} & $<2.8\times10^{46}$ & $<1.7\times10^{47}$ & $<1.5\times10^{46}$ & $<8.6\times10^{46}$ & $<1.1\times10^{46}$ & $<4.3\times10^{46}$ & $<7.9\times10^{45}$ & $<2.5\times10^{46}$ \\ 
\noalign{\smallskip} 
\object{SN2011ke} & $<5.7\times10^{44}$ & $<3.0\times10^{45}$ & $<2.9\times10^{44}$ & $<1.4\times10^{45}$ & $<1.9\times10^{44}$ & $<7.3\times10^{44}$ & $<7.2\times10^{43}$ & $<3.5\times10^{44}$ \\ 
\noalign{\smallskip} 
\object{PTF11dsf} & $<1.6\times10^{45}$ & $<1.0\times10^{46}$ & $<7.8\times10^{44}$ & $<5.2\times10^{45}$ & $<7.6\times10^{44}$ & $<2.7\times10^{45}$ & $<5.5\times10^{44}$ & $<1.4\times10^{45}$ \\ 
\noalign{\smallskip} 
\object{PS1-11afv} & $<5.1\times10^{46}$ & $<3.4\times10^{47}$ & $<3.2\times10^{46}$ & $<1.8\times10^{47}$ & $<2.2\times10^{46}$ & $<9.1\times10^{46}$ & $<9.5\times10^{45}$ & $<4.4\times10^{46}$ \\ 
\noalign{\smallskip} 
\object{SN2011kf} & $<5.7\times10^{44}$ & $<5.3\times10^{45}$ & $<3.9\times10^{44}$ & $<2.7\times10^{45}$ & $<2.5\times10^{44}$ & $<1.4\times10^{45}$ & $<1.8\times10^{44}$ & $<7.2\times10^{44}$ \\ 
\noalign{\smallskip} 
\object{PTF11rks} & $<3.4\times10^{44}$ & $<2.5\times10^{45}$ & $<2.5\times10^{44}$ & $<1.2\times10^{45}$ & $<1.7\times10^{44}$ & $<8.0\times10^{44}$ & $<1.0\times10^{44}$ & $<3.2\times10^{44}$ \\ 
\noalign{\smallskip} 
\object{SN2012il} & $<2.3\times10^{44}$ & $<5.8\times10^{45}$ & $<1.8\times10^{44}$ & $<2.7\times10^{45}$ & $<1.0\times10^{44}$ & $<1.2\times10^{45}$ & $<6.6\times10^{43}$ & $<5.9\times10^{44}$ \\ 
\noalign{\smallskip} 
\object{PTF12dam} & $<6.4\times10^{43}$ & $<7.0\times10^{44}$ & $<4.6\times10^{43}$ & $<3.6\times10^{44}$ & $<3.8\times10^{43}$ & $<1.8\times10^{44}$ & $<3.2\times10^{43}$ & $<9.8\times10^{43}$ \\ 
\noalign{\smallskip} 
\object{LSQ12dlf} & $<5.5\times10^{44}$ & $<9.1\times10^{45}$ & $<3.6\times10^{44}$ & $<4.8\times10^{45}$ & $<2.8\times10^{44}$ & $<2.5\times10^{45}$ & $<1.4\times10^{44}$ & $<2.8\times10^{45}$ \\ 
\noalign{\smallskip} 
\object{SSS120810} & $<1.7\times10^{44}$ & $<1.6\times10^{45}$ & $<1.2\times10^{44}$ & $<8.1\times10^{44}$ & $<7.9\times10^{43}$ & $<4.2\times10^{44}$ & $<7.5\times10^{43}$ & $<3.1\times10^{45}$ \\ 
\noalign{\smallskip} 
\object{CSS121015} & $<1.4\times10^{45}$ & $<6.2\times10^{45}$ & $<8.8\times10^{44}$ & $<3.2\times10^{45}$ & $<4.4\times10^{44}$ & $<2.3\times10^{45}$ & $<3.0\times10^{44}$ & $<1.2\times10^{45}$ \\ 
\noalign{\smallskip} 
\object{iPTF13ajg} & $<1.0\times10^{46}$ & $<5.1\times10^{46}$ & $<7.4\times10^{45}$ & $<3.0\times10^{46}$ & $<4.4\times10^{45}$ & $<1.4\times10^{46}$ & $<3.1\times10^{45}$ & $<7.5\times10^{45}$ \\ 
\noalign{\smallskip} 
\object{SN2013dg} & $<7.1\times10^{44}$ & $<4.7\times10^{45}$ & $<5.6\times10^{44}$ & $<2.6\times10^{45}$ & $<3.9\times10^{44}$ & $<1.2\times10^{45}$ & $<2.7\times10^{44}$ & $<6.2\times10^{44}$ \\ 
\noalign{\smallskip} 
\object{SN2013fc} & $<3.0\times10^{42}$ & $<1.7\times10^{43}$ & $<1.9\times10^{42}$ & $<8.6\times10^{42}$ & $<1.5\times10^{42}$ & $<4.9\times10^{42}$ & $<1.2\times10^{42}$ & $<2.7\times10^{42}$ \\ 
\noalign{\smallskip} 
\object{DES13S2cmm} & $<1.1\times10^{46}$ & $<6.1\times10^{46}$ & $<5.7\times10^{45}$ & $<3.2\times10^{46}$ & $<2.7\times10^{45}$ & $<1.5\times10^{46}$ & $<2.2\times10^{45}$ & $<8.0\times10^{45}$ \\ 
\noalign{\smallskip} 
\object{SN2013hx} & $<2.0\times10^{44}$ & $<1.3\times10^{45}$ & $<8.7\times10^{43}$ & $<6.5\times10^{44}$ & $<7.4\times10^{43}$ & $<4.4\times10^{44}$ & - & - \\ 
\noalign{\smallskip} 
\object{LSQ14an} & $<4.3\times10^{44}$ & $<1.3\times10^{45}$ & $<3.5\times10^{44}$ & $<7.3\times10^{44}$ & $<2.0\times10^{44}$ & $<6.8\times10^{44}$ & - & - \\ 
\noalign{\smallskip} 
\object{iPTF13ehe} & $<1.5\times10^{45}$ & $<5.5\times10^{45}$ & $<7.7\times10^{44}$ & $<2.7\times10^{45}$ & $<4.8\times10^{44}$ & $<1.4\times10^{45}$ & - & - \\ 
\noalign{\smallskip} 
\object{LSQ14mo} & $<8.2\times10^{44}$ & $<8.6\times10^{45}$ & $<5.6\times10^{44}$ & $<3.2\times10^{45}$ & $<3.9\times10^{44}$ & $<1.8\times10^{45}$ & - & - \\ 
\noalign{\smallskip} 
\object{CSS140222} & $<9.5\times10^{42}$ & $<8.0\times10^{43}$ & $<5.9\times10^{42}$ & $<4.1\times10^{43}$ & $<3.1\times10^{42}$ & $<2.0\times10^{43}$ & - & - \\ 
\noalign{\smallskip} 
\object{LSQ14bdq} & $<2.1\times10^{45}$ & $<1.8\times10^{46}$ & $<1.5\times10^{45}$ & $<9.0\times10^{45}$ & $<9.7\times10^{44}$ & $<2.8\times10^{45}$ & - & - \\ 
\noalign{\smallskip} 
\object{PS1-14bj} & $<2.8\times10^{45}$ & $<2.4\times10^{46}$ & $<3.8\times10^{45}$ & $<1.4\times10^{46}$ & $<2.1\times10^{45}$ & $<6.9\times10^{45}$ & - & - \\ 
\noalign{\smallskip} 
\object{DES14X2byo} & $<1.4\times10^{46}$ & $<1.4\times10^{47}$ & $<8.9\times10^{45}$ & $<6.3\times10^{46}$ & - & - & - & - \\ 
\noalign{\smallskip} 
\object{DES14X3taz} & $<5.7\times10^{45}$ & $<3.6\times10^{46}$ & $<3.9\times10^{45}$ & $<1.8\times10^{46}$ & - & - & - & - \\ 
\noalign{\smallskip} 
\object{LSQ15abl} & $<6.7\times10^{43}$ & $<6.6\times10^{44}$ & $<6.1\times10^{43}$ & $<3.2\times10^{44}$ & - & - & - & - \\ 
\noalign{\smallskip} 
\object{SN2015bn} & $<2.4\times10^{44}$ & $<2.0\times10^{45}$ & $<1.3\times10^{44}$ & $<4.6\times10^{44}$ & - & - & - & - \\ 
\noalign{\smallskip} 
\object{ASASSN-15lh} & $<5.0\times10^{44}$ & $<3.5\times10^{45}$ & - & - & - & - & - & - \\ 
\noalign{\smallskip} 
\hline 
%\vspace{1.5cm}
\end{tabular}}
\end{table} 

\end{appendix}

\end{document}